\documentclass[twocolumn]{aastex63}
\usepackage{amsmath}
\usepackage{ mathrsfs }

\hypersetup{linkcolor=blue,citecolor=cyan,filecolor=green,urlcolor=magenta}

\received{19 June 2020}
\revised{xxx}
\accepted{xxx}

\shorttitle{OGLE Reanalysis}
\shortauthors{Golovich et al.}

\begin{document}
\title{A Reanalysis of Public Galactic Bulge Gravitational Microlensing Events from OGLE-III and IV}

\author[0000-0003-2632-572X]{Nathan Golovich}
\added{\correspondingauthor{Nathan Golovich}
\email{golovich1@llnl.gov}}
\affiliation{Lawrence Livermore National Laboratory, 7000 East Avenue, Livermore, CA 94550, USA}

\author[0000-0003-0248-6123]{William Dawson}
\affiliation{Lawrence Livermore National Laboratory, 7000 East Avenue, Livermore, CA 94550, USA}

\author[0000-0001-8630-9794]{Fran Bartoli\'{c}}
\affiliation{Centre for Exoplanet Science, School of Physics and Astronomy, University of St. Andrews, St. Andrews KY16 9SS, UK.}
\affiliation{Center for Computational Astrophysics, Flatiron Institute, 162 5th Avenue, 6th Floor, New York, NY 10010, USA }

\author[0000-0002-6406-1924]{Casey Y. Lam}
\affiliation{University of California, Berkeley, Department of Astronomy, Berkeley, CA 94720, USA}

\author[0000-0001-9611-0009]{Jessica R. Lu}
\affiliation{University of California, Berkeley, Department of Astronomy, Berkeley, CA 94720, USA}

\author[0000-0002-7226-0659]{Michael S. Medford}
\affiliation{University of California, Berkeley, Department of Astronomy, Berkeley, CA 94720, USA}
\affiliation{Lawrence Berkeley National Laboratory, 1 Cyclotron Rd., Berkeley, CA 94720, USA}

\author[0000-0002-8505-7094]{Michael D. Schneider}
\affiliation{Lawrence Livermore National Laboratory, 7000 East Avenue, Livermore, CA 94550, USA}

\author{George Chapline}
\affiliation{Lawrence Livermore National Laboratory, 7000 East Avenue, Livermore, CA 94550, USA}

\author[0000-0002-3569-7421]{Edward F. Schlafly}
\affiliation{Lawrence Livermore National Laboratory, 7000 East Avenue, Livermore, CA 94550, USA}

\author[0000-0001-8251-933X]{Alex Drlica-Wagner}
\affiliation{Wilson Fellow, Fermi National Accelerator Laboratory, Kirk Road and Pine St, Batavia, IL 60510, USA}
\affiliation{Department of Astronomy and Astrophysics, University of Chicago, 5640 South Ellis Avenue, Chicago, IL 60637, USA}

\author[0000-0002-2911-8657]{Kerianne Pruett}
\affiliation{Lawrence Livermore National Laboratory, 7000 East Avenue, Livermore, CA 94550, USA}

\begin{abstract}
Modern surveys of gravitational microlensing events have progressed to detecting thousands per year, and surveys are capable of probing Galactic structure, stellar evolution, lens populations, black hole physics, and the nature of dark matter. One of the key avenues for doing this is the microlensing Einstein radius crossing time ($t_E$) distribution.
However, systematics in individual light curves as well as over-simplistic modeling can lead to biased results.
To address this, we developed a model to simultaneously handle microlensing parallax due to Earth’s motion, systematic instrumental effects, and unlensed stellar variability with a Gaussian Process model.
We used light curves for nearly 10,000 OGLE-III and IV Milky Way bulge microlensing events and fit each with our model. We also developed a forward model approach to infer the $t_E$ distribution by forward modeling from the data rather than using point estimates from individual events. 
We find that modeling the variability in the baseline removes a source of significant bias in individual events and previous analyses overestimated the number of $t_E>100$ day events due to their over-simplistic model ignoring parallax effects. We use our fits to identify hundreds of events that have a black hole lens purity of roughly $50\%$. Finally, we have released the largest ever catalog of Markov-chain Monte Carlo parameter estimates for microlensing events.
\end{abstract}


\section{Introduction} \label{sec:intro}
Gravitational microlensing was first proposed as a means to study the dark halo of the Milky Way galaxy by \citet{1986ApJ...304....1P}, and has since yielded direct constraints on the population of dark objects \citep{1996ApJ...471..774A, 1996astro.ph.12102E, 1998ApJ...499L...9A, 2001ApJ...550L.169A, 2002astro.ph.12176E, 2007AA...469..387T}. These were achieved through time-domain photometric surveys of the densest stellar regions of the night sky (the Milky Way bulge and Magellanic clouds) with optical telescopes \citep{1992AcA....42..253U, 1993ASPC...43..291A, 1994ExA.....4..265A, 1999PThPS.133..233M}. The field of microlensing has pushed dense field photometry, difference imaging photometry, transient alerts, and many more technologies that will be used by upcoming surveys (e.g., the Legacy Survey of Space and Time -- LSST and the Wide Field Infrared Space Telescope -- WFIRST) for a variety of science goals. Though the heritage of microlensing was in the identification of dark objects to understand the nature of dark matter, most practitioners are on the hunt for exoplanets today \citep[see][for a review and a projection, respectively]{2012ARAA..50..411G, 2019ApJS..241....3P}.

Setting binary systems and exoplanets aside, modeling even a simple point-source, point-lens (PSPL) microlensing event is challenging. First, given that most surveys point toward extremely dense stellar regions, it is virtually impossible to measure the flux of the source and lens alone for most events. Given typical seeing from the ground, the baseline flux of most microlensing events are likely composed of several stars' light \citep{2019ApJ...871..205S}, and given the frequency at which stars exhibit intrinsic stellar variability, the baseline is likely to be variable itself. On top of this, since the Galactic bulge is near the ecliptic, it is near the Sun periodically. The seasonal gaps in the light curve lead to aliasing and the variations of atmospheric seeing, weather patterns, and other systematics must be considered, especially for long duration events which span many of these cycles and seasonal gaps. Additionally, the motion of the Earth around the Sun complicates the modeling. The Earth's orbital parallax effect (caused by variable magnification due to Earth's motion around the Sun) is also on a yearly timescale, so care must be taken to not allow this physical signal to mix with the systematics.

Most Galactic bulge microlensing events are short (measurable magnification for at most a couple months) such that the Earth's motion can be either outright ignored or assumed as a constant acceleration. A number of degeneracies arise with the latter assumption \citep[see e.g.,][]{2004ApJ...606..319G}. However, microlensing parallax measurements are useful given they help constrain the mass-distance degeneracy \citep{2016MNRAS.458.3012W}. Microlensing parallax is also useful for exoplanet studies allowing numerous degeneracies in binary lens caustics to be resolved \citep{2015ApJ...802...76Y}. Given the usefulness, microlensing parallax is frequently modeled; however, given the potential for correlations with troublesome systematics, there has been no systematic study of microlensing parallax across entire surveys.

As microlensing surveys began detecting upwards of thousands of microlensing events, distributions of microlensing parameters were studied, especially the global $t_E$ (Einstein crossing time) distributions. The $t_E$ distribution has become an important proxy for physical inference on the Galactic structure. These studies probed several details regarding the population of lenses -- including at the short $t_E$ end, the existence of free-floating planets \citep{2011Natur.473..349S, 2017Natur.548..183M}. The global shape of the distribution has been studied to compare observations \citep{2015ApJS..216...12W,  2019ApJS..244...29M} to simulations that probe the present-day mass function of lenses \citep{1996ApJ...473...57M}. The shape of the distribution has also been used to probe the matter profile in the inner galaxy \citep{1996ApJ...473...57M, 2016MNRAS.463..557W, 2017ApJ...843L...5W, 2020ApJ...889...31L}. \citet{2015ApJS..216...12W} used it to probe the spatial distribution of the optical depth in the Galactic bulge region. \citet{2019ApJS..244...29M} used it to show the optical depth and microlensing event rate over roughly 40 square degrees with higher spatial resolution and showed that the bulge event rates compared well with the expected microlensing event rate predicted by a Besançon stellar model \citep[see, e.g.,][]{2014AA...564A.102C}. In general; however, the $t_E$ distribution is dependent on the line of sight distribution of source and lens parallax and relative proper motion \citep{2019ApJS..244...29M}.

The $t_E$ distribution is also a rough proxy of the mass distribution of lenses \citep[see e.g.,][]{1996ApJ...473...57M}, and can be used to constrain the fraction of intermediate mass ($\sim10-1000 M_\odot)$ primordial black hole dark matter \citep{1992ApJ...392..442G, 2019RNAAS...3...58L}. \citet{2019RNAAS...3...58L} showed that OGLE could begin to probe this with the $t_E$ distribution; however, \citet{2020ApJ...889...31L} showed in a more thorough analysis of \textsf{PopSyCLE} simulations that utilizing parallax and astrometric measurements would greatly increase the purity of black hole populations from microlensing. We will explore this question in more detail in two follow on studies (Dawson et al. in prep, Pruett et al. in prep).

Black holes in microlensing surveys need not be primordial in nature. The Milky Way has $10^8-10^9$ black holes of stellar origin \citep{2002MNRAS.334..553A}. The exact number is uncertain, and methods to detect them, other than microlensing, are limited, with most incapable of detecting isolated black holes \citep[i.e., those not in binaries;][]{2006ARAA..44...49R}. Given the clear existence of black holes in the $\sim 5-30$ M$_\odot$ mass range from binary detections, it is reasonable to expect that isolated black holes have likely been detected already in microlensing surveys. The typical time-scale of these events would be $\sim 100$ days\footnote{See below for a full discussion of the microlensing math, but assuming 4 kpc to the lens, 8 kpc to the source, and 200 km s$^{-1}$ transverse velocity of the lens to the line of sight, 5-30 M$_\odot$ gives lensing event durations of $\sim78-191$ days.}, and events with this duration have been observed frequently. The problem is that estimating the mass from the photometric signal that microlensing surveys probe is challenging. Assuming the same source distance, the mass--distance degeneracy ensures that a high-mass and far away lens can provide a similar observed lensing effect to a low-mass and nearby lens. This degeneracy is broken with the parallax-effect due to the Earth's orbit (which causes asymmetric magnification as well as ``wiggles'' in the light curve) and with astrometric observations of the lensing behavior (which causes unresolved multiple images of the source and appears as a deviation from the linear proper motion of the flux centroid). Parallax is notoriously challenging to model and the systematic observation of all events with astrometry is not possible given the time-consuming observations required. There is also a degeneracy between massive fast moving lenses and low mass, slow moving lenses that is difficult to disambiguate without high resolution follow up over years.

If the route of carefully modeling the Earth's orbital parallax is taken, then additional challenges arise. Black holes are more massive than typical stars, so the relative parallax between the source and lens is more likely to be small. Small parallax manifests less obvious signal -- rather than the obvious periodic ``wiggles'' in the photometric light curve of a long duration microlensing event, the rise and fall of the microlensing signal is very slightly asymmetric. When we consider that microlensing signal in seeing-limited surveys is typically highly-blended with unlensed flux  due to numerous neighbor stars and consider the frequency that stars exhibit measureable photometeric variability, there is often a time-varying signal due to baseline variability in either the source or blended flux. This signal makes constraining the parallax challenging when it is intrinsically small ($\pi_E<0.1$, see \S\ref{sec:event_model} for definition of $\pi_E$). The timescale of the microlensing event (the Einstein crossing time, $t_E$) is also useful to analyze in the context of black holes, as it contains information about the mass of the lens and distances to and relative proper motion of the lens and source. Individual events with long timescales have been scrutinized as black hole candidates \citep{2002ApJ...579..639B, 2002ApJ...576L.131A,2002MNRAS.329..349M,2005ApJ...633..914P}. In principle long duration and low parallax microlensing events are most likely to be black holes \citep{2020ApJ...889...31L}; however, given the systematics described above, picking out and constraining the parallax to a small value with the signal correlated over several observing seasons is a challenge.

Gaussian process (GP) models are often used for problems like these. This additional time-domain correlated noise has been fit with GP models across time-domain astronomy including transiting exoplanet observations with Kepler \citep[see e.g.;][]{2009MNRAS.395.2226B, 2017AJ....154..220F}. This method has only very recently been used in microlensing studies of an individual event \citep{2019MNRAS.488.3308L}, and has not been widely deployed on an entire survey. This is unsurprising given the computational requirements to invert large matrices repeatedly in order to infer the hyper-parameters of the GP kernel.

A recent theoretical study of OGLE-like microlensing survey simulations using a powerful new tool suggests that many microlensing events are indeed black holes \citep[PopSyCLE;][]{2020ApJ...889...31L}; however, given the aforementioned challenges it seems a population approach is needed. We undertook the considerable task of analyzing all public Galactic bulge OGLE-III and IV microlensing light curves and estimating posteriors for the microlensing model parameters (including parallax) as well as inferring GP hyper-parameters in order to subtract away the time-domain correlated noise that afflicts proper inference on the microlensing parameters. In \S\ref{sec:ogle}, we describe the input data for our individual light curve modeling. In \S\ref{sec:event_model}, we describe the model and sampling algorithm to estimate the posterior probability density function (PDF) for all of the model parameters. In \S\ref{sec:pop_model}, we describe a powerful method of combining all of the information from our parameter posterior samples and demonstrate a useful analysis on the Einstein crossing time global distribution. In \S\ref{sec:results}, we present our individual light curve results as well as our results from the global Einstein crossing time analysis before we discuss and conclude our study in \S\ref{sec:discussion}. We are releasing our MCMC parameter estimate chains for all 9350 light curves analyzed. These will be available online at XXX. 

Throughout this paper, all timescale parameters have the units of days (measured in modified Julian days) and flux quantities assume $m_{I}=22$ for a zero-point in the OGLE published magnitude system. We will discuss throughout three different microlensing models. The math for each of these will be developed in \S\ref{sec:ogle} and \S\ref{sec:event_model}. First is the standard five-parameter \citet{1996ARAA..34..419P} model, which assumes a point-source and point-lens (PSPL) with no parallax and a constant baseline with blended, unlensed flux. We will refer to this model as PSPL throughout. Second is the parallax model, still assuming PSPL but with the added physics of Earth's motion. We will refer to this as PSPL$+$parallax model. Third is the most complicated model, which includes both parallax and a Gaussian Process model for the variability in the baseline. We will refer to this as the PSPL$+$parallax$+$GP model. Finally, we will discuss a PSPL$+$GP model, but only for plotting purposes in \S\ref{subsec:results_indiv}. 

\section{Data} \label{sec:ogle}
\subsection{OGLE-III}\label{subsec:OGLEIII}
The OGLE-III survey spanned the years of 2002--2009 and was operated from Las Campanas Observatory in Chile on the 1.3 meter Warsaw University Telescope. The observations were collected with eight (4$\times$2) 2048$\times$4096 CCD chips with 0$\arcsec$.26 pixel$^{-1}$ with a total field of view of $35\arcmin \times 35\arcmin$. For more details, we refer readers to the survey paper \citep{2003AcA....53..291U}. OGLE-III published alerts online\footnote{\url{http://ogle.astrouw.edu.pl/ogle4/ews/ews.html}} for each of their eight seasons with 4000 events on their Early Warning System. OGLE-III began the era of statistical inference on populations of thousands of microlensing events. \citet{2015ApJS..216...12W} published 3718 unique microlensing events, which was at the time was the largest ever catalog of identified microlensing events. In their analysis, each event was fit with a  the I-band flux as:
\begin{equation}
I=I_0-2.5\,\log\left(b_{\text{sff}} A + \left( 1 - (b_{\text{sff}} \right)\right)
\end{equation}where $(b_{\text{sff}}$ is the source flux fraction or fraction of the total baseline flux by the source (note \citet{2015ApJS..216...12W} labels this $f_S$, but we've adopted the notation of \citet{2020ApJ...889...31L} to avoid confusion with the flux of the source star);
\begin{equation}\label{eq:mag}
    A=\frac{u^2+2}{u\sqrt{u^2+4}}
\end{equation}is the amplification of the baseline source flux, and
\begin{equation}
    u=\sqrt{u_0^2+\frac{\left(t-t_0\right)^2}{t_E^2}}
\end{equation}is the projected angular separation between the lens and source in units of the Einstein angle $\theta_E$:
\begin{equation}
    \theta_E = \sqrt{\frac{4GM}{c^2}\frac{D_\text{S}-D_\text{L}}{D_\text{L}D_\text{S}}},
\end{equation}which is dependent on the lens mass ($M$) and distances to the lens ($D_\text{L}$) and source ($D_\text{S}$); $I_0$ is the baseline magnitude of the light curve in the OGLE $I$-band. It is convenient to define
\begin{equation}
    \tau=\frac{t-t_0}{t_E}.
\end{equation}where 
 and $t_0$ and $u_0$ are defined as $u\left(t_0\right)\equiv u_0$ in the heliocentric frame, and $t_E$ is the Einstein crossing time, or the time it takes the source to traverse the lens' angular Einstein radius:
\begin{equation}\label{eq:eq15}
    t_E=\frac{\theta_E}{\mu_{\text{rel}}},
\end{equation} where $\mu_{\text{rel}}$ is the projected lens--source relative proper motion. The modeled parameters are then 
\begin{equation}
    \boldsymbol{\theta}_\text{PSPL} = \{t_0,t_E,u_0, I_0,f_S\}.
\end{equation}\citet{2015ApJS..216...12W} first fit this model to all light curves using a $\chi^2$ minimization code in order to use these values as features in a random-forest classifier for event detection along with many other light curve characteristics. There was also an upper limit of $t_E=400$ days applied to all the OGLE-III data. The full OGLE-III catalog was then reduced to 3560 events, which were then modeled with a Markov Chain Monte Carlo (MCMC) analysis, which estimates the posterior distributions for the parameters described above. 

In this paper, we will work with the 3560 $I$-band light curves as published in \citet{2015ApJS..216...12W}, perform new fits, and we will draw comparisons to their analysis of the distribution of fit parameters of their events. 

\subsection{OGLE-IV}\label{subsec:OGLEIV}
The OGLE-IV survey commenced in 2010 and continues today with a 32 chip (2048$\times$4096 pixel) mosaic camera with 0$\arcsec$.26 pixel$^{-1}$ with a total field of view of 1.4 square degrees. The upgraded camera also had a much shorter readout time allowing for an order of magnitude increase in survey efficiency over OGLE-III. The most recent full-scale analysis of OGLE-IV was presented in \citet{2019ApJS..244...29M}, which analyzes eight years of OGLE-IV data and amasses a catalog of 8000 microlensing events toward the Galactic bulge fields of the OGLE-IV survey. This increase in statistical power allowed for higher resolution modeling of Galactic structure toward the bulge and higher fidelity population statistics of microlensing parameters. Similar to the OGLE-III analysis \citep{2015ApJS..216...12W}, this study published the individual light curves for microlensing events they detected, as well as \citet{1996ARAA..34..419P} curve fits to each event. Similarly to \citet{2015ApJS..216...12W}, they also imposed an upper limit on the time-scale ($t_E<300$ days). 

We will use the $I$-band light curves as published in our analysis to follow. We note here that this is not a comprehensive list of OGLE-IV events. It is missing the central-most Galactic fields, which were published separately in \citet{2017Natur.548..183M} as a stand-alone analysis of the high cadence light curves in search for free-floating planets. These 2617 light curves were not made public, and so they are left out of our study. Additionally we did not include 630 OGLE-IV events in the Galactic plane recently made public and analyzed in \citet{2020arXiv200407289M}.

\subsection{PopSyCLE Simulation}

A third set of microlensing events that we will use for comparisons to guide analysis are from a recent simulation that serves as a detailed mock OGLE type survey \citep{2020ApJ...889...31L}. For details on the simulation and recording of microlensing events we refer readers to \citet{2020ApJ...889...31L}, but in short, \textsf{PopSyCLE} (Population Synthesis for Compact object Lensing Events) is an open source tool for simulating microlensing surveys starting with a stellar model of the Milky Way, evolving stars through stellar evolution models across an initial-final mass relation, stochastically assigning compact object populations to stars that evolve to that stage, and simulating a microlensing survey on the final Galactic catalog. We make use of the mock OGLE-III and IV simulations presented in \citet{2020ApJ...889...31L} for comparisons to our analysis (v2 -- see their Appendix A).

\section{Event Modeling}\label{sec:event_model}

In this section, we describe the modeling framework that we will carry out in order to fit all OGLE-III and OGLE-IV events described in \S\ref{sec:ogle}.

\subsection{Annual Parallax}

In \S\ref{subsec:OGLEIII} we described the basic \citet{1996ARAA..34..419P} microlensing model used by both \citet{2015ApJS..216...12W} and \citet{2019ApJS..244...29M} to model individual events. This model is an approximation that ignores the motion of the Earth around the Sun and models the measured flux as the sum of an unlensed and lensed flux as well as additional complexities such as binary sources and/or lenses, and finite source effects \citep[see e.g.,][]{1998AA...329..361D}.

In this subsection, we introduce the effect of observing microlensing from the Earth. Long $t_E$ events or good black hole candidates are sensitive to parallax effects. This is coupled with multiple degeneracies in the model space; i.e., multiple combinations of parameters can yield approximately the same light curve to within the modeling precision. These degeneracies arise from numerous situations including blending of lensed and unlensed light \citep{1997ApJ...487...55W, 2009MNRAS.393..816D}, the geometry of gravitational lensing generically \citep[the so-called mass--distance degeneracy; see e.g.,][]{2000AJ....120.1654S}, and the fact that \citet{1996ARAA..34..419P} parameters are heliocentric in nature but estimated in the geocentric frame \citep{2003MNRAS.339..925S,2004ApJ...606..319G}. The last of these, observing distant lenses and sources from a moving reference frame reference frame, especially when the timescale of the microlensing event is long, is particularly challenging. 

The annual parallax was first observed in a microlensing event by the MACHO collaboration \citep{1995ApJ...454L.125A}. Our formulation for microlensing parallax is based on a combination of \citet{1995ApJ...454L.125A} and  \citet{2013ApJ...763L..35G}. We first reconciled Eq. 2 from \citet{1995ApJ...454L.125A} with Eq. 4 from \citet{2013ApJ...763L..35G} in order to add the eccentricity of Earth's orbit to the formalism of \citet{2013ApJ...763L..35G}\footnote{
In our reconciliation of \citet{1995ApJ...454L.125A} and \citet{2013ApJ...763L..35G} we discovered that Eq.\ 2 from \citet{1995ApJ...454L.125A} is missing an additive term $2\tau u_0 \sin\left(2\phi\right).$
}. We then modeled the parallax as follows. 

\begin{figure}
\plotone{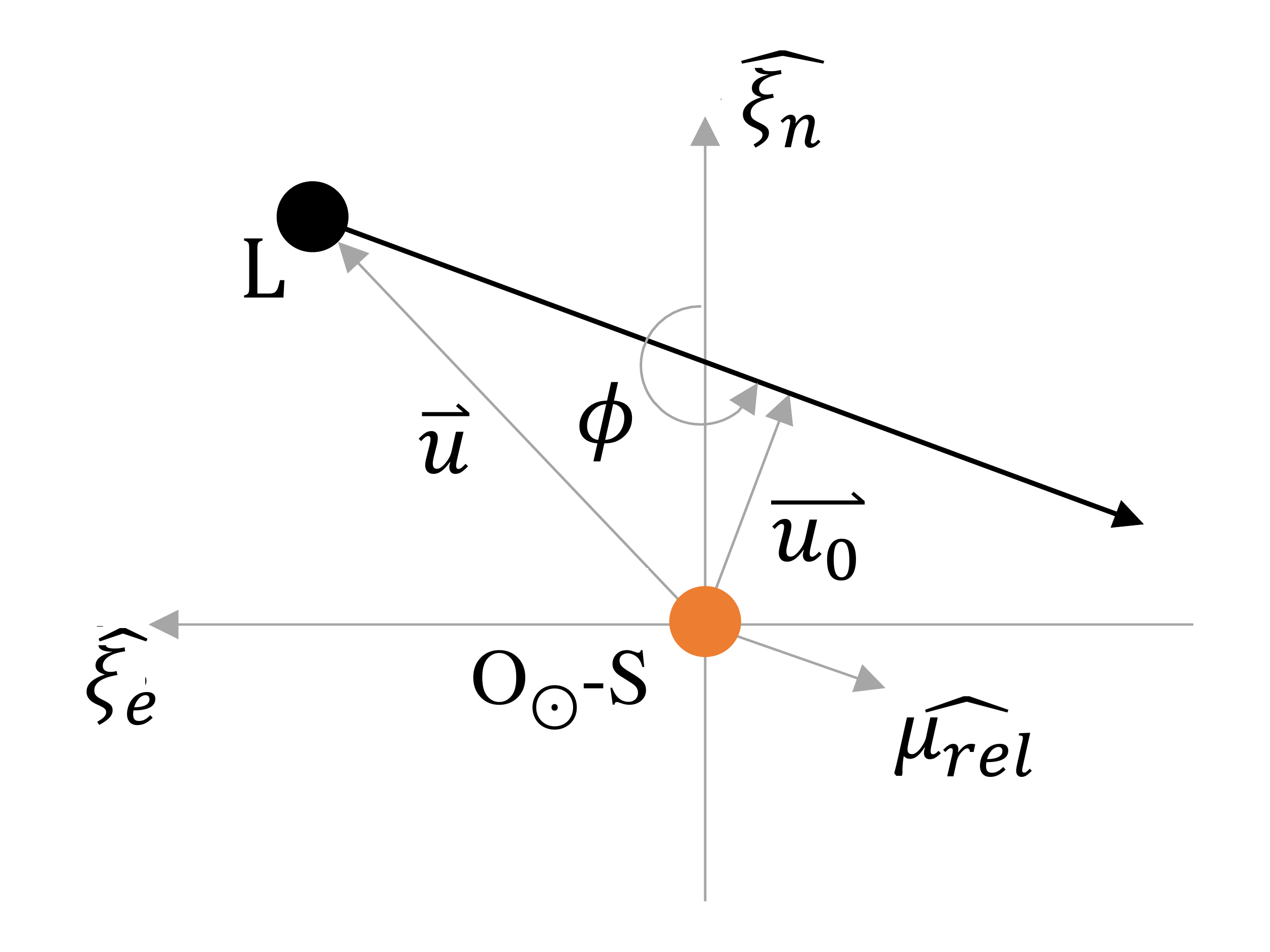}
\caption{The geometry in the plane of the sky normal to the heliocentric observer-source axis. The lens (black circle) passes to the north and west of the source with a relative proper motion direction characterized by the angle $\phi$ with a minimum angular separation of $u_0$. These are each model parameters in our analysis. The unit vectors $\hat{\xi}_n$ and $\hat{\xi}_e$ are aligned with the ecliptic coordinate latitude and longitude, respectively. The unit vector $\hat{\mu}_{rel}$ is in the direction of the relative motion of the lens with respect to the $O_\odot$-S coordinate. \label{fig:fig1}}
\end{figure}

\begin{figure}
\plotone{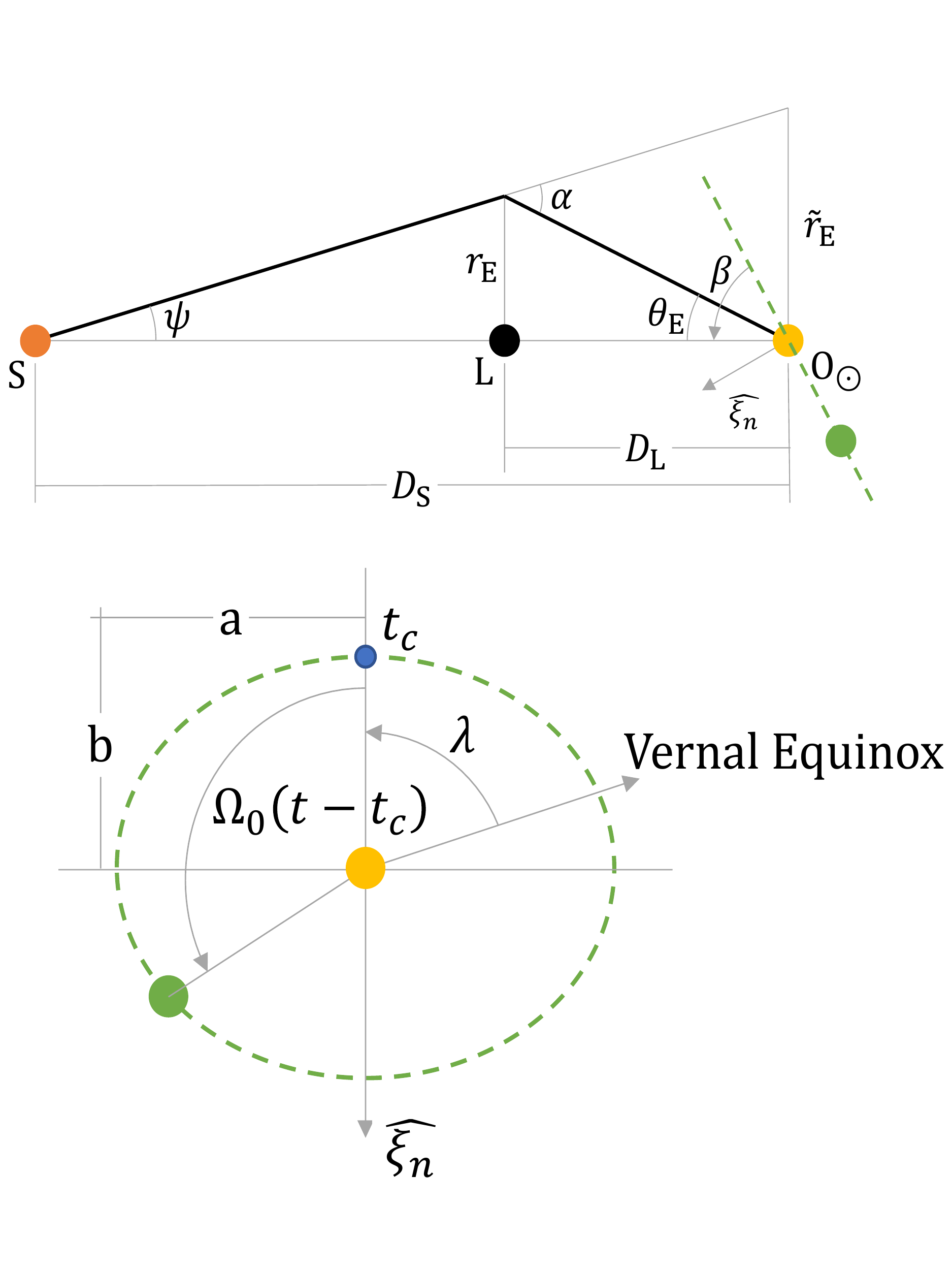}
\caption{\emph{Top:} The microlensing formalism of \citet{2000ApJ...542..785G} augmented by microlensing parallax relevant parameters. The bold line represents the path of light from the source (S) to the heliocentric observer ($O_\odot$), which is deflected by the lens (L). $D_\mathrm{L}$ and $D_\mathrm{S}$ are the assumed Euclidean distances to the lens and source from $O_\odot$, respectively. The deflection angle $\alpha$ is a function of the Einstein radius $r_E\,(\alpha = 4GM/r_E c^2)$, while $\theta_E$ is the Einstein angular radius on the sky. The Einstein radius projected onto the S-O normal plane is defined as $\Tilde{r}_E$. In this projection, Earth's orbit (green dashed line) is angled by the ecliptic latitude of the source star $\beta$. \emph{Bottom:} The orbit of Earth (dashed green ellipse) projected onto the S-$O_\odot$ normal plane. Note that the ellipticity, and corresponding semi-major and semi-minor axes $a$ and $b$, respectively, are representative of a nearly circular orbit projected on the plane. The frame is orientated such that the time, $t_c$, at which Earth is closest to the $O_\odot$-S line is at the top. Within the orbital plane of Earth, $\lambda$ is the angle of $t_c$ from the vernal equinox. Note that $t_c$ can be determined by the time of vernal equinox, $t_{VE}$ and the ecliptic longitude of S, $\lambda\,\,(t_c = \frac{\lambda}{2\pi}T\,+t_{VE})$, where $T$ is the Earth's orbital period in days. The angle $\beta$ in the left figure is simply the ecliptic latitude of S and $\xi_n$ is in the plane of the page and corresponds to the northern celestial coordinate unit vector. The phase angle of Earth at time $t$ is given by $\lambda + \Omega_0 * (t-t_c)$, where $\Omega_0=2\pi/T$ is the conversion factor for the orbital time to radians.\label{fig:fig2}}
\end{figure}

The observed data of microlensing is a flux time-series or ``light curve'' that is typically constructed either from point spread function (PSF) fit photometry or differential imaging analysis (DIA). Most generally, the flux in the light curve is the sum of all sources that lie along the line of sight inside of the extraction profile. However, since the projected angular extent of the detectable lensing effect is much smaller than the typical PSF full width at half maximum (FWHM), it is difficult to determine which of these sources is the one being lensed. We separate the flux in the light curve into two categories: $S$ for the source, $B$ for the blended unlensed flux (this includes lens flux). Then the flux may be modeled as:

\begin{equation}\label{eq:flux}
    F\left(t\right)=F_B\left(t\right)+F_S\left(t\right)\,A\left(t\right)
\end{equation} where the individual, unlensed fluxes are time-varying due to stellar variability in general, and from General Relativity \citep{1936Sci....84..506E}, $A\left(t\right)$ is given by Eq. \ref{eq:mag}; $u$ is still the magnitude of the lens--source separation in the plane normal to the observer--source axis in units of the Einstein angle $\theta_E$, 
\begin{equation}\label{eq:eq6}
    \theta_E = \sqrt{\frac{4GM}{c^2}\left(\frac{1}{D_L}-\frac{1}{D_S}\right)},
\end{equation}
but it is modified by Earth's motion.
Note that $M$ is the lens mass and $D_L$ and $D_S$ are the lens and source distance, respectively. $G$ is Newton's gravitational constant, and $c$ is the speed of light. Accounting for Earth's motion, the lens--source separation in geocentric coordinates, $\mathbf{u}$, is a vector with ecliptic components given by:
\begin{multline}\label{eq:u}
    \mathbf{u}= \left[\tau \sin{\phi} + u_0 \cos{\phi} + \xi_e \, \pi_E \sin{\Omega}\right] \hat{\xi}_e  \\ + \left[\tau \cos{\phi}-u_0 \sin{\phi}+\xi_n \, \pi_E \cos{\Omega} \right] \hat{\xi}_n,
\end{multline} 
see Figure \ref{fig:fig1}. For circular orbits,
\begin{equation}
    \xi_e=\frac{a}{AU} \equiv \xi ; \,\,\, \xi_n = \frac{b}{AU}= \xi \sin\beta,
\end{equation}
where $a$ is the semi-major axis of Earth's orbit, $b$ is the semi-minor axis of the circular orbit's projected ellipse on the plane perpendicular to the $O_\odot$-S axis, and $\beta$ is the ecliptic latitude, see Figure \ref{fig:fig2}. $\phi$ is the angle between the Ecliptic North and the direction of the source-lens relative proper motion in the heliocentric frame. Accounting for the eccentricity, $\epsilon$, of Earth's orbit generalizes this to:
\begin{equation}
    \xi_e=\xi\,\left[1-\epsilon \cos{\left[\Omega_0\left(t-t_p\right)\right]}\right]
\end{equation}
\begin{equation}
    \xi_n=\xi \sin{\beta} \left[1-\epsilon \cos{\left[\Omega_0\left(t-t_p\right)\right]}\right],
\end{equation}where $t_p$ is the time of the perihelion and $\Omega_0$ is the angular frequency of Earth's orbit:
\begin{equation}
    \Omega_0=\frac{2\pi}{T},
\end{equation} where $T$ is the period of the Earth's orbit around the Sun;
\begin{equation}
    \Omega = \Omega_0\left(t-t_c\right) + 2\,\epsilon\,\sin{\left[\Omega_0\left(t-t_p\right)\right]},
\end{equation} where $t_c$ is the time of the Earth's closest approach to the Sun--source axis:
\begin{equation}
    t_c = \frac{\lambda}{2\pi} T + t_{VE},
\end{equation} where $\lambda$ is the equatorial longitude of the source and $t_{VE}$ is the time of the vernal equinox, see bottom panel of Figure \ref{fig:fig2}.
Finally, the microlensing parallax $\pi_E$ is
\begin{equation}
    \pi_E = \frac{\pi_{\text{rel}}}{\theta_E},
\end{equation} where $\pi_{\text{rel}}$ is the relative lens--source parallax.

\subsection{Intrinsic Variability of the Baseline}
In most microlensing analyses to date it is assumed that $F_B$, $F_S$ and $F_L$ are constant in time thus rendering the baseline ($A\left(t\right) \rightarrow 1$) flat at large $|t-t_0|$, with the exception of \citet{2019MNRAS.488.3308L}, who modeled the baseline variability of a single event using a GP model. This assumption is never perfect as all stars are variable at some level due to the complex physics of each star and all light curves are afflicted with seasonal throughput variations due to weather and other systematics; however, it may be a reasonable assumption for the average microlensing event that has $t_E\sim20$\,days. Since we are just as interested in long timescale events, where stellar variability is more likely to be a significant systematic, we desire a generic model that can fit all OGLE-III and OGLE-IV events. 

To model the time-variability of the baseline of the light curve, we use a GP model. For a good summary of GP models and their use in statistical modeling and machine learning we refer readers to \citet{RasmussenWilliams}. In short, a GP is a probability distribution over possible functions. This is an important aspect that allows for the use of Bayes' Theorem and thus integrate the formalism into our preferred probabilistic sampling algorithms. More concretely, for the problem of modeling a microlensing light curve, we use a GP model to handle the variable baseline and correlated noise in the light curve. This is necessary because of the myriad of sources of systematic noise in our observed data (e.g., detector peculiarities, stellar variability, weather, sky conditions, among countless others). The use of a GP may be thought of (and appears in our plots) as fitting the residual of the physical model, but we caution against this view point. In actuality, we are simultaneously fitting both the GP model and the physical model parameters, i.e.:
\begin{equation}
    F\left(t\right) = \text{physics}\left(t\right)+\text{GP}\left(t\right)+\text{white\,noise}. 
\end{equation}

Conditioned on a finite number of observations, a GP is just a multivariate Gaussian distribution with a dense covariance matrix. We must define the $\boldsymbol{\mu}$ and $\boldsymbol{\Sigma}$ --- mean and covariance matrix, respectively --- for this multi-variate Gaussian to set up such a distribution. GP models are simplest if we center the data (i.e., allow $\mu=0$). In principle we want to allow for relationships between points in our light curve (this is simple to imagine in two dimensions. The covariance matrix is generated by evaluating the kernel function $\kappa$, which takes two points $t_i$ and $t_j$ and returns the similarity measure between them as a scalar:
\begin{multline}
        \kappa: \mathbb{R}^n \times \mathbb{R}^n \rightarrow \mathbb{R};\; \Sigma_{ij}  = \kappa\left(\tau_{ij}\right)+K^2\,\sigma_i^2\,\delta_{ij}
\end{multline} where $\tau_{ij}=|t_i-t_j|$, $\sigma_i$ is the typical uncertainty in the $i$-th data point, and $\delta_{ij}$ is the Kronecker-$\delta$. We have also included a scaling factor, $K$, on the uncertainty of each measurement which helps handle under-estimated uncertainty in flux measurements. This means that choosing a kernel function strongly influences on the range of shapes the function can take. We refer readers to \citet{Duvenaud} for an overview of constructing kernels. 

Computationally, adding a GP to our likelihood function means we now must invert and compute the determinant of a covariance matrix. To avoid an unsophisticated $\mathcal{O}\left(N^3\right)$ implementation (for $N$ flux measurements in the light curve) we make use of the \textsf{celerite} python package \citep{2017AJ....154..220F}, which scales as $\mathcal{O}\left(N\right)$ through a number of simplifications on the kernel choices that are physically motivated for stellar flux time-series measurements. 

The variability in the baseline is complex. It is due to a number of causes including variations of photometric quality over the course of the year and asteroseismic oscillations and other internal phenomena that occur at timescales related to the internal structure of the stellar system but are also damped by dissipation. For this reason we use a damped simple harmonic oscillator (SHO) term for the kernel provided by \textsf{celerite}. \citet{2017AJ....154..220F} shows the underlying kernels that are motivated by the Fourier transform solutions to a damped, stochastically-driven, damped SHO (the derivation proceeds through their Eq. 19--31). The natural kernels that come from this physical picture are shown to be quasi-periodic and described with exponential kernels, which \textsf{celerite} exploits to vastly speed up the evaluation. The power spectral density of this kernel is given by:
\begin{equation}
    S\left(\omega\right) = \sqrt{\frac{2}{\pi}}\frac{S_0\,\omega_0^4}{\left(\omega^2-\omega_0^2\right)^2+\omega_0^2\,\omega^2/Q^2}
\end{equation} where $\omega_0$ is the undamped frequency, $Q$ is the quality factor of the oscillator, and $S_0$ a constant proportional to the undamped spectral power. We set $Q=1/\sqrt{2}$ since leaving it unconstrained gave the GP model unlimited freedom to fit the microlensing event itself. This also dramatically improves the speed of hyper-parameter estimation and reduces the kernel to:

\begin{multline}
        \kappa_{\text{SHO}}\left(\tau_{ij}\right) = S_0\,\omega_0\,\exp{\left(-\frac{1}{\sqrt{2}}\,\omega_0\,\tau_{ij}\right)} \\ \times \cos{\left(\frac{\omega_0\,\tau_{ij}}{\sqrt{2}}-\frac{\pi}{4}\right)}.
\end{multline}

This kernel has been used for background granulation noise in time-series flux measurements of stars \citep{1985ESASP.235..199H, 2009AA...495..979M, 2014AA...570A..41K, 2017AJ....154..220F}. The source of this noise has been described as a combination of several effects including convection, acoustic oscillations, magnetic activity, and rotation \citep{2014ApJ...781..124C}; these are all studied in the rich literature of astroseismology \citep[see][for a comprehensive review]{2010aste.book.....A}. 

\begin{figure}
    \centering
    \includegraphics[width=\columnwidth]{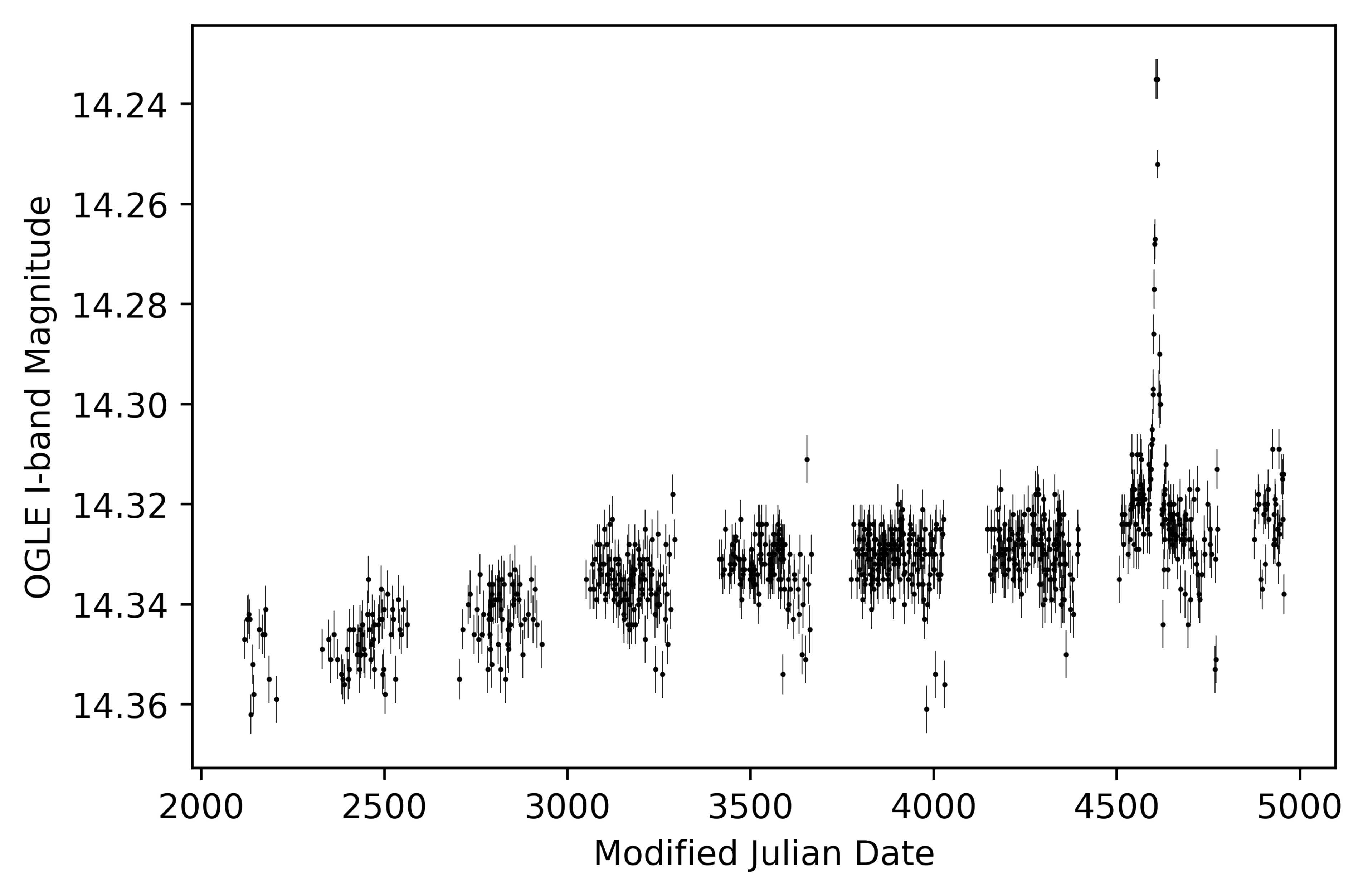}
    \caption{Example OGLE-III light curve (OGLE BLG 156.7.141434) showing the baseline of the light curve. The baseline is clearly rising over the course of several years in addition to a quasi-periodic oscillatory behavior. We have designed our GP kernel to handle these systematics specifically since these are frequent in OGLE data.}
    \label{fig:fig3}
\end{figure}

In Figure \ref{fig:fig3} we show a portion of an example light curve from OGLE-III (see \S\ref{subsec:OGLEIII}). There is clearly a quasi-periodic baseline in the wing of the microlensing event. The SHO kernel term should handle the periodic fluctuations, but we must also handle the general trend of the baseline. This low-order polynomial trend is extremely troublesome for sampling the posterior of the microlensing parameters of interest (e.g., $\pi_E$ and $t_E$). Without handling the time-varying baseline, the model is forced into extremely unlikely regions of parameter space in order to explain both the true underlying microlensing event which varies over a small fraction of the whole light curve and the nearly linear trend over the several years that flux measurements were made. To handle this type of light curve, we add a Mat\'{e}rn-$3/2$ kernel. Note that we may construct kernels as the sum or product of two valid kernels \citep{2011arXiv1103.4023D}. The Mat\'{e}rn-$3/2$ kernel is given by:

\begin{equation}
    \kappa_{\text{M3/2}}\left(\tau_{ij}\right) = \sigma^2\left(1+\frac{\sqrt{3}\,\tau_{ij}}{\rho}\right) \exp \left(-\frac{\sqrt{3}\,\tau_{ij}}{\rho}\right),
\end{equation}where $\rho$ and $\sigma$ are the modeled hyper-parameters. 

All together, our data is modeled with a covariance structure according to:
\begin{equation}
    \Sigma_{ij}  = \kappa_\text{SHO}\left(\tau_{ij}\right)+\kappa_\text{M3/2}\left(\tau_{ij}\right)+K^2\,\sigma_i^2\,\delta_{ij},
\end{equation}and the model flux vector is:
\begin{equation}
    \boldsymbol{\mu} = \boldsymbol{\Phi}\,\boldsymbol{w}
\end{equation} where $\boldsymbol{\Phi}$ is:
\begin{equation}
    \boldsymbol{\Phi} = \begin{pmatrix} 
    A\left(t_1\right) & 1 \\
    A\left(t_2\right) & 1 \\
    \vdots & \vdots \\
    A\left(t_N\right) & 1  
    \end{pmatrix}
\end{equation} for $N$ flux measurements, where $A$ is given in Eq. \ref{eq:mag}, and 
\begin{equation}
    \boldsymbol{w} = \begin{pmatrix} 
    F_S \\
    F_B \\
    \end{pmatrix}
\end{equation} where $F_S$ and $F_B$ are the flux of the source and blended light (including any lens flux), respectively. It is convenient to define the baseline, unlensed flux:
\begin{equation}
    F_{\text{base}}=F_S+F_B
\end{equation} and the source-flux-fraction:
\begin{equation}
    b_{\text{SFF}} = \frac{F_S}{F_S+F_B},
\end{equation} which we will sample in our modeling and can be used to select dark lenses \citep{2016MNRAS.458.3012W,2020AA...636A..20W}.

Thus, for observed flux data $\mathbf{F}$, the above formulation allows for a Gaussian likelihood function with mean $\mathbf{\mu}$ and covariance matrix $\boldsymbol{\Sigma}$:
\begin{equation}\label{eq:likelihood}
    p\left(\mathbf{F}|\boldsymbol{\theta}\right) \sim \mathcal{N}\left(\mathbf{F};\boldsymbol{\mu},\boldsymbol{\Sigma}\right),
\end{equation}where $\mathcal{N}$ indicates a normal distribution, and $\boldsymbol{\theta}$ is the set of microlensing and GP hyper-parameters:
\begin{equation}\label{eq:params}
    \boldsymbol{\theta} = \{F_\text{base},b_{\text{SFF}},t_0,t_E,u_0,\pi_E,\phi,S_0,\omega_0,\rho,\sigma,K\}.
\end{equation}

\subsection{Priors}\label{subsec:priors}
In addition to the likelihood of the previous section, we must also define priors on our model parameters in order to properly sample the posterior PDF. We discuss these priors in this subsection individually; however, the key information is contained in Table \ref{tab:priors}.

\begin{table*}
    \begin{center}
    \caption{Summary of prior probability density functions for modeled parameters}\label{tab:priors}
    \begin{tabular}{lcl}
    Parameter (1)   &   Distribution (2)    &   Inputs (3)\\
    \hline
    $F_\text{base}$                     &   $\mathcal{N}$    &   $\mu=\text{med}\left(\boldsymbol{F}\right)$, $\sigma=\sigma_{\boldsymbol{F}}$ \\
    $b_\text{SFF}$                      &   $\mathcal{U}$    &   lower$=0$, upper$=1+\epsilon_\text{NB}$\\
    $t_0$                               &   $\mathcal{N}$    &   $\mu=t_{0,\text{OGLE}}$, $\sigma=100$ days\\
    $\log_{10}t_E$                      &   $\mathcal{N}$    &   $\mu=1.13435$, $\sigma = 0.67502$\footnote{based on PopSyCLE output}, truncated between $0.5<t_E<3000$ days \\
    $\log_{10}u_0$                      &   $\mathcal{N}$    &   $\mu=-1$, $\sigma=2$, truncated between $10^{-5}<u_0<3$\\
    $\log_{10}\pi_E$                    &   $\mathcal{N}$    &   $\mu=-0.884205$, $\sigma = 0.072755$\footnote{based on PopSyCLE output}, truncated between $10^{-5}<\pi_E<3$\\
    $\phi$                              &   $\mathcal{U}$    &   lower$=0$, upper$=2\pi$\\
    $\log\left(S_{0}\,\omega_{0}^{4}\right)$   &   $\mathcal{N}$    &  $\mu=\sigma^2_{\boldsymbol{F}}$, $\sigma=5$\\
    $\log \omega_0$                     &   $\mathcal{N}$    &  $\mu=0$, $\sigma = 5$\\
    $\log \sigma$                       &   $\mathcal{N}$    &   $\mu=0$, $\sigma = 5$\\
    $\rho$                              &   $\Gamma^{-1}$    &   see \S\ref{sssec:inv_gamma} for details \\ 
    $\log\left(K^2\right)$              &   $\mathcal{N}$    &   $\mu=0.0953$, $\sigma = 1$\\
    \end{tabular}
    \end{center}

    \tablecomments{Column 1 gives the parameters in our PSPL$+$parallax$+$GP model. Column 2 gives the distribution assumed for the prior PDF for each parameter. $\mathcal{N}$, $\mathcal{U}$, and $\Gamma^{-1}$ indicate a normal, uniform, and inverse gamma distribution, respectively. Col. 3 gives the input values for each of these distributions. For the normal distributions, a mean ($\mu$) and standard deviation ($\sigma$) are listed. In some cases these normal distributions are truncated, with zero probability outside the specified range. In the case of uniform distributions, the lower and upper limits are listed, outside this range the prior PDF is zero. We refer readers to the corresponding text for details of the inverse-gamma distribution.}
\end{table*}

\subsubsection{Flux Parameters: $F_{\text{base}}$ and $b_{\text{SFF}}$}
The measured flux of a typical microlensing event is an unknown combination of source, lens, and blended light. It is possible to disambiguate the three using a combination of high resolution follow up and parallax measurements, but in principle we do not know flux ratios a priori. Modeling the flux parameters is a challenge since these are highly correlated with several other microlensing parameters. We thus use a Markov Chain Monte Carlo to sample in the total baseline flux and the source flux fraction with the following priors:
\begin{equation}
    p\left(F_\text{base}\right) \propto \mathcal{N}\left(\text{median}\left(\mathbf{F}\right),\sigma_\mathbf{F}\right),
\end{equation}where the median is taken over the flux time-series and $\sigma_\mathbf{F}$ refers to the standard deviation of the flux time series;
\begin{equation}
    p\left(b_{sff}\right) \propto \mathcal{U}\left(0,1+\epsilon_{NB}\right)
\end{equation} where $\mathcal{U}(a,b)$ describes a uniform distribution between $a$ and $b$ that is zero elsewhere; $\epsilon_{NB}$ allows for a small amount of ``negative blending'' \citep{2004ApJ...609..166P, 2007MNRAS.380..805S}, where the source apparently contributes more than 100$\%$ of the observed flux. This is, strictly speaking, nonphysical, but it becomes possible in the parameters due to systematics in the modeling of the pixels; e.g., if the difference imaging procedure improperly attributes flux from a neighbor to the source. We restrict the amount of negative blending to the amount of flux from a star of $m_I=20.5$. This follows the analysis of \citet{2019ApJS..244...29M}, where a comparison of OGLE fields was made to high resolution \emph{Hubble Space Telescope} images and pixel-level simulations (see Figures 8 and 9 of \citet{2019ApJS..244...29M}). 
\subsubsection{Time of Closest Approach: $t_0$}
The time of closest approach is defined in the heliocentric frame; i.e, it is the time when, as viewed from the Sun, the lens and source are nearest in their trajectories on the sky. This is troublesome for modeling light curves with large microlensing parallax signals because then the peak of the light curve is different than $t_0$, thus we cannot set a tight prior on $t_0$ to the peak of the light curve. Since high microlensing parallax is rare, and most light curves have good agreement between $t_0$ and the peak of the light curve, we use a Gaussian prior, centered on the reported $t_0$ values from OGLE \citep[][]{2015ApJS..216...12W,2019ApJS..244...29M};
\begin{equation}
    p\left(t_0\right) \propto \mathcal{N}\left(t_{0,\text{OGLE}},100\,\text{days}\right).
\end{equation}
\subsubsection{Einstein Crossing Time: $t_E$}
The time it takes for the source to transit the angular Einstein radius, $\theta_E$, is again defined in the heliocentric reference frame is parametrized by $t_E$. \citet{2009MNRAS.393..816D} showed reparametrizations that are useful for fitting Earth-based microlensing observations including defining the effective-timescale $t_\text{eff} = t_E\,u_0$ and modeling in $u_0$ and $t_\text{eff}$. We use $t_E$ and $u_0$ parameters in our analysis since $t_\text{eff}$ is most useful for shorter events, and we want to be capable over a large range in $t_E$. The time-scale distribution of microlensing surveys is often reported as a histogram of $t_E$ point estimates from the population of events. For OGLE-III and IV, this distribution has proven to have a peak roughly at 20 days with tails toward short timescales as short as 1 day and as long as several hundred days. Recently, \citet{2020ApJ...889...31L} published the first simulated OGLE-like microlensing survey which largely confirmed these findings (though the peak of the $t_E$ distribution is slightly lower). We use the output from the \citet{2020ApJ...889...31L} OGLE-like \textsf{PopSyCLE} simulation to develop our $t_E$ prior. However, we inflate the distribution to allow for the possibility of longer timescale events in the OGLE database. We sample in $\log_{10}{t_E}$, so our $t_E$ prior is given by: 
\begin{equation}\label{eq:tE}
    p\left(\log_{10}t_E\right) \propto \mathcal{N}\left(\mu_{\log_{10} t_E}, 2\sigma_{\log_{10} t_E}\right)
\end{equation} where the mean and standard deviation refer to the values from the \textsf{PopSyCLE} OGLE simulation ($\mu_{\log_{10}} t_E = 1.13435$ and $\sigma_{\log_{10}} t_E = 0.33751$. In addition, we truncate this prior to only allow for $0.5<t_E<3000$ days. 
\subsubsection{Impact Parameter: $u_0$}
The impact parameter is defined as $u_0
\equiv u\left(t_0\right)$. It is again defined in the heliocentric frame and often reparametrized as a factor of the effective timescale ($t_\text{eff}=t_E\,u_0$). Geometrically, given random sources and lenses with random motion on the sky, the distribution of impact parameters for any pair is roughly uniform out to large radii. However, when modeling a population of microlensing events, this is weighted by the detection sensitivity of the survey. Large impact parameter microlensing events are difficult to detect since the peak magnification gets buried in the noise of the flux measurements. We sample in $\log_{10}{u_0}$ under the prior:
\begin{equation}
    p\left(\log_{10}u_0\right) \propto \mathcal{N}\left(-1, 2\right)
\end{equation} truncated between $10^{-5}<u_0<3$. Note that we are only modeling the positive $u_0$ solutions. We discuss this choice in more detail in \S\ref{sec:pop_model}. We checked that there is no bias in the population modeling made by this choice.

\subsubsection{Microlensing Parallax, $\pi_E$}
The microlensing parallax is a measure of the effect of the Earth's orbit relative to the size of the projected angular Einstein radius. No population distribution of this parameter has ever been constructed from data, so we again turn to the \textsf{PopSyCLE} simulation results. We sample in $\log_{10} \pi_E$ under the prior:
\begin{equation}
    p\left(\log_{10}\pi_E\right) \propto \mathcal{N}\left(\mu_{\log_{10} \pi_E}, \sigma_{\log_{10} \pi_E}\right)
\end{equation} truncated between $10^{-5}<\pi_E<3$. Here, again, the mean and standard deviation values are computed from the \textsf{PopSyCLE} output ($\mu_{\log_{10} \pi_E} = -0.884205$ and $\sigma_{\log_{10} \pi_E} = 0.072755$). 

\subsubsection{Relative Proper Motion Angle: $\phi$}
In most characterizations of microlensing parallax, it is treated as a vector quantity with magnitude equal to the relative parallax between the lens and source in units of the angular Einstein radius. The direction of the vector is, by convention, aligned with the lens--source relative motion vector and often quoted in ecliptic coordinate components. In our parametrization we instead model the magnitude of the parallax and, separately, the angle $\phi$ measured from the Ecliptic North to the direction of the projected relative proper motion of the source and lens. This proper motion is in the heliocentric coordinate frame. We sample under a uniform parallax angle prior:
\begin{equation}
    p\left(\phi\right) \propto \mathcal{U}\left(0,2\pi\right).
\end{equation} In practice, sampling in a periodic angle can cause issues since the sampler sees a discontinuity, so we sampled in $\sin{\phi}$ and $\cos{\phi}$ following the method developed in the \textsf{exoplanet} python package \citep{exoplanet}.

\subsubsection{Simple Harmonic Oscillator Gaussian Process Kernel Priors: $S_0$ and $\omega_0$}
Our model includes a stochastically driven, damped SHO covariance function as part of the covariance matrix for the likelihood function. This GP term has two hyper-parameters $S_0$ and $\omega_0$. These are related to the power spectral density at the characteristic frequency of the SHO described by the kernel. They are often highly correlated, and since the power-spectral density of this kernel is a function of $S_0\,\omega_0^4$ we sample in $\log{\left(S_0\,\omega_0^4\right)}$ and $\log \omega_0$ with priors:
\begin{equation}
    p\left(\log \left(S_0\,\omega_0^4\right)\right) \propto \mathcal{N}\left(\sigma_F^2,5\right),
\end{equation}
\begin{equation}
    p\left(\log \left(\omega_0\right)\right) \propto \mathcal{N}\left(0,5\right),
\end{equation}where $\sigma_F^2$ is the variance of the flux time-series, following the \textsf{celerite} tutorials for time series data.

\subsubsection{Mat\'{e}rn-3/2 Gaussian Process Kernel Priors: $\sigma$ and $\rho$}\label{sssec:inv_gamma}
Our model also includes a Mat\'{e}rn-$3/2$ kernel term, which is parametrized by $\rho$ and $\sigma$, which are the characteristic time-scale and amplitude of the flux variability, respectively. We allow for a wide range of flux variability:
\begin{equation}
    p\left(\log \left(\sigma\right)\right) \propto \mathcal{N}\left(0,4\right).
\end{equation}
For the time-scale hyper-parameter, we use an inverse-gamma prior with 1\% of probability below the typical data spacing and above the length of the time-series \footnote{Following \url{https://betanalpha.github.io/assets/case_studies/gp_part1/part1.html}}: 
\begin{equation}
    p\left(\rho\right) \propto \Gamma^{-1}\left( a,b \right).
\end{equation}

\subsubsection{Diagonal Covariance Scale Factor: $K$}
Finally, we model an additional parameter which linearly scales the uncertainty of the flux measurements. This is sampled following the prior:
\begin{equation}
    p\left(\log{\left(K^2\right)}\right) \propto \mathcal{N}\left(\log\left(1.1\right),1\right).
\end{equation}
\subsection{Sampling}
Our model has twelve parameters and many known degeneracies and correlations between subsets of the model parameters. Thus, in order to sample from the posteriors we must make use of an efficient sampling algorithm. In the previous section we listed one by one the prior PDFs we will use. The likelihood function is given in Eq. \ref{eq:likelihood}. Each likelihood evaluation is expensive given the GP in our model. We make use of \textsf{celerite} and \textsf{exoplanet} packages to not only  speed up this portion of our analysis, but also because these packages were built into the framework of \textsf{PyMC3} \citep{2015arXiv150708050S}, which is a Python package for Bayesian statistical modeling. We make use of the No-U-Turn Sampler \citep[NUTS;][]{2011arXiv1111.4246H}, which is related to the Hamiltonian Monte Carlo (HMC) algorithm. HMC is a MCMC algorithm that uses gradient information to converge with complicated posteriors more quickly than Metropolis or Gibbs sampling \citep[for a review, see][]{2017arXiv170102434B}. NUTS is an improvement on HMC because it computes important tuning parameters as it samples and removes the potential for user error. The NUTS implementation in \textsf{exoplanet} computes the mass matrix internally through a series of tuning steps that enable more efficient sampling of the dense regions of high-dimensional posterior PDFs \citep{exoplanet}. 

After we tune the sampler with 4000 tuning steps, we sample all 3560 OGLE-III light curves and 5790 OGLE-IV light curves under the described model for 40,000 samples (20,000 steps of burn-in and 20,000 saved samples). We provide the chains for all 9350 posterior samples for all parameters in an online data set totaling roughly 23 gigabytes. 

\section{Population Modeling}\label{sec:pop_model}
In the microlensing literature, many authors present distributions of point estimates over each of their fits. This is especially prevalent with the $t_E$ \citep{2011Natur.473..349S,2015ApJS..216...12W,2017Natur.548..183M,2019ApJS..244...29M}. Given that microlensing modeling is riddled with degeneracies and correlations, there is reason to believe these distributions are potentially biased given that they were modeled without parallax and the variability of the baseline was not modeled. In this section, we will discuss proper Bayesian methods for combining the full posteriors from each light curve.

\subsection{A Hierarchical Model}
Often, MCMC methods are used to sample the complicated posteriors for binary lens and source systems throughout the recent literature, and there are now open-source modeling packages available to the community, e.g., \textsf{pyLIMA} \citep{2018AAS...23122805B} and \textsf{MulensModel} \citep{2019A&C....26...35P}.

As Bayesian methods have become more popular, authors have provided a varying amount of this information in the literature. Some authors do not report posteriors and only the maximum likelihood, MAP or the marginal median values for each parameter along with typical confidence intervals on modeled parameters (often 68\%, 90\% or 95\% credible intervals). Some others provide graphical representations including corner plots, which show the marginal two-dimensional posteriors and correlations in addition to the marginal one-dimensional distributions. Others still provide all samples. We have chosen to provide all samples, so our posteriors can be maximally useful for comparisons and extensions to our work as well as any subsequent analysis.

In this subsection, we will demonstrate a use case with the $t_E$ chains. We follow the method described by \citet{2010ApJ...725.2166H}, which develops the probability theory for combining samples of eccentricity posteriors from the analysis of individual exoplanet orbits from radial velocity data. We refer readers to that paper for the full details. In short the method sets out to forward model the distribution of the parameter of interest itself, rather than estimate the distribution from histograms of single point estimates or summed MCMC samples which can result in biased estimates. 

We have $N$ microlensing events labeled $n\,\left(1\leq n \leq N\right)$ with $M_n$ flux measurements at times $t_j$, labeled $j\,\left(1\leq j \leq M_n\right))$, $F_{nj}$. From these data and the likelihood function given in Eq.\ \ref{eq:likelihood} and the prior PDFs  $\left(p\left(\boldsymbol{\theta_n}\right)\right)$ in \S\ref{subsec:priors} we have computed samples from the the individual posterior PDFs:
\begin{equation}
    p\left(\boldsymbol{\theta_n}|F_n\right) = \frac{p\left(F_n|\theta_n\right)p\left(\boldsymbol{\theta_n}\right)}{Z_n},
\end{equation} where $Z_n$ is a normalization constant. Then, assuming all of the light curves are independent (note this is mostly true, but sources of telescope and detector systematic error make it not fully true) the total likelihood for all of the light curves parameters is the product of each likelihood function. Instead we want the likelihood for the set of parameters, $\boldsymbol{\alpha}$, that describe the distribution of interest over the individual chains, $f_{\boldsymbol{\alpha}}\left(t_{E}\right)$. We can think of this distribution as a better prior PDF for the parameter of interest. Mathematically this amounts to a change of variables and integrating out the individual light curve parameters. Following \citet{2010ApJ...725.2166H}:
\begin{equation}
\begin{split}
    \mathscr{L}_{\boldsymbol{\alpha}} & \equiv p\left(\{F_n\}^N_{n=1}|\boldsymbol{\alpha}\right) \\
    & = \prod_{n=1}^N \int d\boldsymbol{\theta}_n\, p\left(F_n|\boldsymbol{\theta}_n\right) p\left(\boldsymbol{\theta}_n|\boldsymbol{\alpha}\right),
\end{split}
\end{equation}where,
\begin{equation}
    p\left(\boldsymbol{\theta}_n|\boldsymbol{\alpha}\right) = \frac{f_{\boldsymbol{\alpha}}\left(t_E\right)\,p\left(\boldsymbol{\theta}_n\right)}{p\left(t_E\right)}.
\end{equation}

The integrals are taken over $12N$ dimensions, and yield a marginal likelihood or probability of the data given the parameters $\boldsymbol{\alpha}$. This integral is not practical in most cases, but since we have already generated K-element samplings with elements $\boldsymbol{\theta}_{nk}$:
\begin{equation}\label{eq:integrals}
    \int d\boldsymbol{\theta}_{n}\,p\left(\boldsymbol{\theta}_{n}|F_{n}\right)f\left(\boldsymbol{\theta}_n\right) \approx \frac{1}{K}\sum_{k=1}^{K} f\left(\boldsymbol{\theta}_{nk}\right),
\end{equation}where the posterior PDF in the integrand represents the posteriors from the individual fits, and $f\left(\boldsymbol{\theta}_{n}\right)$ is an arbitrary function of the parameters. Since all of the integrals in Eq.\  \ref{eq:integrals} can be approximated by sums over samples of the individual posteriors, the marginal likelihood for $\boldsymbol{\alpha}$ is:
\begin{equation}
    \mathscr{L}_{\boldsymbol{\alpha}} \approx \prod_{n=1}^{N}\, \frac{1}{K}\,\sum_{k=1}^{K}\,\frac{f_{\boldsymbol{\alpha}}\left(t_{E_{nk}}\right)}{p\left(t_{E_{nk}}\right)}.
\end{equation}Here the sum is over the ratio of the new prior we want to infer over the prior we used for $t_E$ in the individual light curve model. Importantly, we must make sure the old prior has support over the full range of interest for the new distribution. \citet{2010ApJ...725.2166H} recommends uninformative priors for the individual fits; however, as long as there is support over the full domain of $f_{\boldsymbol{\alpha}}\left(t_E\right)$, this need not be entirely uninformative. In fact, uninformative priors are often informative in ways that the modeler did not intend\footnote{See \citet{2015PASP..127..994B} for an illustrative example.}. We note that our prior is supported in the range, 0.5 days $<t_E<$ 3000 days with a smooth normal distribution in $\log_{10}\left(t_E\right)$. With this likelihood function, we can then multiply by a prior PDF for $\boldsymbol{\alpha}$, normalized and sampled to obtain a posterior PDF distribution for $\boldsymbol{\alpha}$ fully forward modeled from the individual light curves. 

Now we turn our attention to choosing a functional form for $f_{\boldsymbol{\alpha}}\left(t_E\right)$ and a prior PDF for $\boldsymbol{\alpha}$. We make relatively simple choices for each of these since our intention is to demonstrate the power of our publishing of chains for all public Galactic bulge OGLE-III and OGLE-IV microlensing events from \citet{2015ApJS..216...12W} and \citet{2019ApJS..244...29M}, respectively. We model the distribution as a step-function following \citet{2010ApJ...725.2166H}. This is useful to compare to back to traditional histograms used in point estimate distributions presented in \citet{2015ApJS..216...12W,2019ApJS..244...29M}. It is sort of non-parameteric model (the parameters are the bin heights themselves) where we don’t have to assume a overly simplified functional form. 
\begin{equation}
    f_{\boldsymbol{\alpha}}\left(\log_{10}\left(t_E\right)\right) \equiv \sum_{m=1}^{M}\,\alpha_m\,s\left(t_E;\,\frac{m-1}{M},\frac{m}{M}\right),
\end{equation}where
\begin{equation}
    s\left(x;\,L,\,H\right) \equiv
        \begin{cases}
          0, & \text{for}\ x<L \\
          (H-L)^{-1}, & \text{for}\ L \leq x \leq H \\
          0, & \text{for}\ H<x
        \end{cases},
\end{equation}and
\begin{equation}
    \sum_{m=1}^M\,\alpha_m=1. \label{eq:condition}
\end{equation}
This can be thought of as a normalized histogram with $M$ equal width bins in $\log_{10}\left(t_E\right)$.
For the prior PDF on $\boldsymbol{\alpha}$ (i.e., the bin amplitudes) we use a Dirichlet prior:
 \begin{equation}
 p\left(\boldsymbol{\alpha}\right) \propto \mathrm{Dir}\left(\boldsymbol{\alpha}|\boldsymbol{a}\right).
 \end{equation}

\begin{figure*}
    \includegraphics[width=\textwidth]{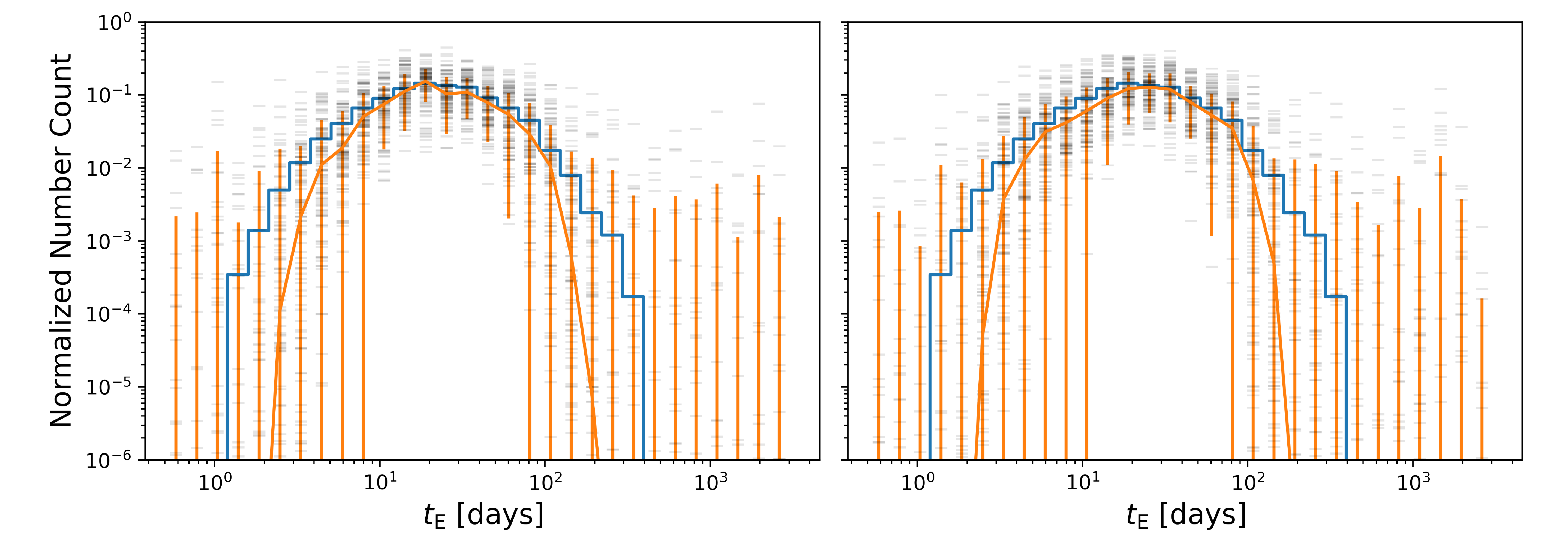}
    \caption{Graphical representation for the prior probability density function for the population modeling of the Einstein crossing time distribution for OGLE-III (left panel) and OGLE-IV (right panel). The blue histogram in each panel shows the normalized bin heights for the MAP $t_E$ values for our individual fits. The orange bars show the 68\% confidence interval, about the mean (solid orange curve), for random samples from the Dirichlet prior distribution (shown in gray ticks). The Dirichlet priors are for $b=20$ and $c=0.02$.}\label{fig:fig4}
\end{figure*}

 Here the Dirichlet distribution is parametrized by the Dirichlet concentration vector $\boldsymbol{a}$, which has the same number of elements $M$ as $\boldsymbol{\alpha}$. The Dirichlet prior describes a vector valued multinomial distribution which has the convenient property in that it naturally satisfies Eq.\ \ref{eq:condition}. We construct $\boldsymbol{a}$ by conditioning on the weights $\left(\boldsymbol{w}\right)$ for a histogram of MAP $t_E$ values for all events, from the respective survey, and compute the Dirichlet concentration parameters as:
 \begin{equation}
     \boldsymbol{a} = \left(\boldsymbol{w} - \text{min}\left(\boldsymbol{w}\right)\right)\,b+c.
 \end{equation}
 $b$ tunes the allowable bin amplitude variance about the prior, where the larger $b$, the less variance there will be for each bin amplitude $m$ away from $w_m$, $c$ influences how uniform the bin height prior is across the $M$ bins, where $c<1$ will more closely follow $\boldsymbol{w}$, and $c>1$ will approach uniform. A full Bayesian analysis would open these up for modeling; however, we tested a large range for both of these parameters and found negligible difference in the results for the $t_E$ distribution. We set $b=20$ and $c=0.02$. The priors for both OGLE-III and OGLE-IV are shown in Figure \ref{fig:fig4}. 
 
\subsection{Sampling}
For sampling this model, we again make use of \textsf{PyMC3} and the NUTS algorithm. We sample the model over all $t_E$ chains for OGLE-III and OGLE-IV separately. We sample for 3200 steps, burning the first 1600. We again ensured that the number of effective samples was sufficiently large to avoid biased results from auto-correlation. The results of this analysis are presented in \S\ref{subsec:results_pop}. 

\section{Results}\label{sec:results}
In \S \ref{sec:event_model} we described in detail the model and sampling algorithm that we used to generate posterior samples for 12 individual event microlensing and GP hyper-parameters simultaneously for 3560 OGLE-III and 5790 OGLE-IV microlensing events. In \S \ref{sec:pop_model}, we described the advantage of modeling the population distribution of important microlensing parameters by importance sampling the individual samples generated in \S\ref{sec:event_model} and estimating distribution parameters in a Bayesian, hierarchical framework. In this section, we will present the key results from these sections separately. 

\subsection{Individual Events}\label{subsec:results_indiv}
Each light curve has been sampled with the likelihood function given in Eq. \ref{eq:likelihood} (the PSPL$+$parallax$+$GP model) and prior PDFs on twelve parameters given in \S\ref{subsec:priors}. We also computed the MAP values for all twelve parameters as well as the representative MAP estimates for the PSPL$+$parallax$+$GP model and the PSPL model. The MAP models were computed with a Broyden–Fletcher–Goldfarb–Shanno (BFGS) algorithm \citep[see e.g.,][]{BFGS}. The posterior PDF for 12 parameters for all of the publicly available OGLE-III \citep{2015ApJS..216...12W} and OGLE-IV \citep{2019ApJS..244...29M} Galactic bulge microlensing events were sampled, and all samples are available for download with this publication. In this subsection we will present the results for a few representative examples.

\begin{figure*}
    \includegraphics[width=\textwidth]{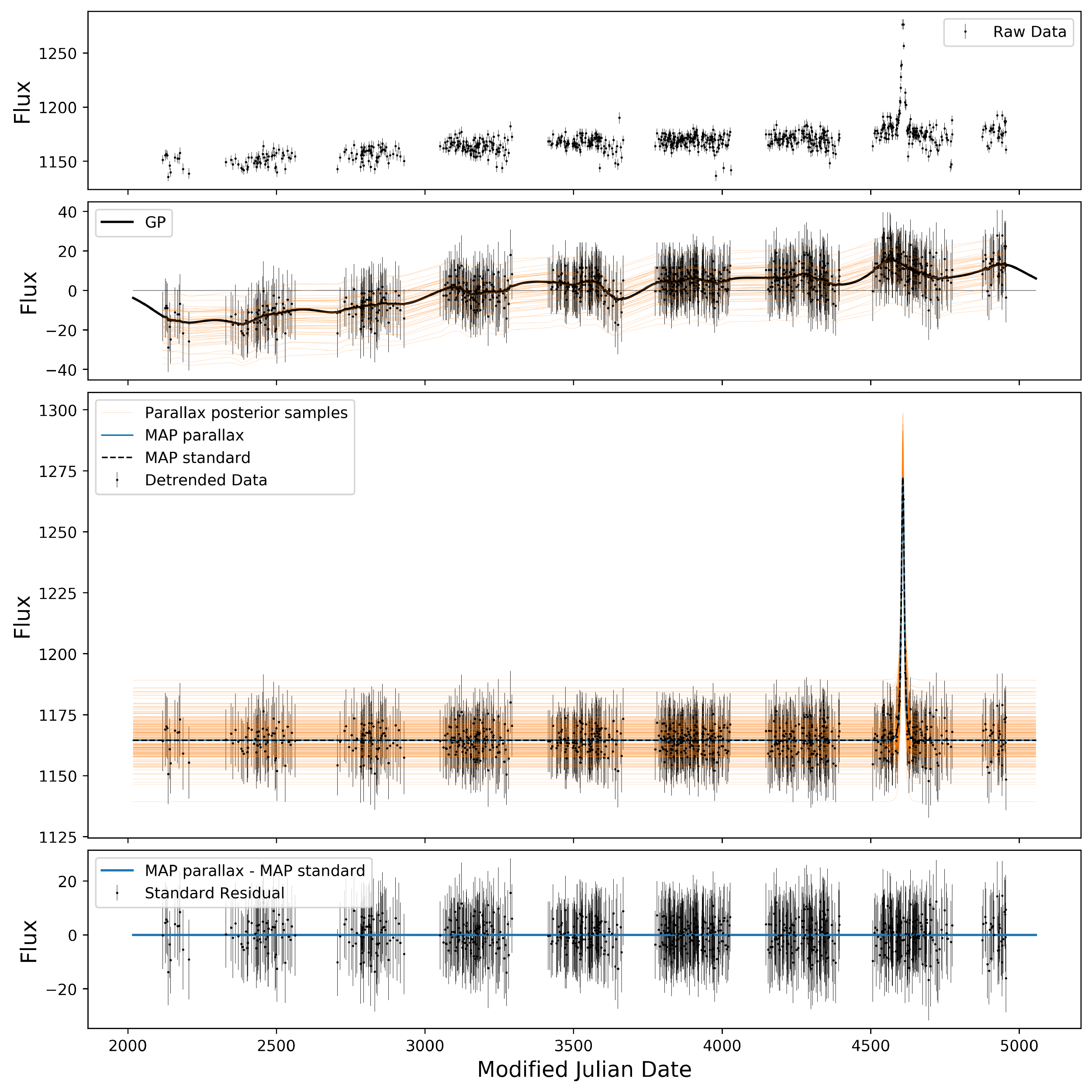}
    \caption{Example OGLE-III light curve (OGLE BLG 156.7.141434) with results from our MCMC analysis. \emph{First panel:} Raw data published by \citet{2015ApJS..216...12W}; \emph{Second panel:} GP component of the PSPL$+$parallax$+$GP model with the MAP in thick black and 50 random draws from our posteriors in orange. The data has the MAP PSPL$+$parallax subtracted from the raw data in the first panel and rescaled uncertainty on each measurement by the MAP value of $K$ in our model. \emph{Third panel:} 200 microlensing posterior samples (orange) and MAP parallax model (solid blue) subtracting away the MAP GP from the panel above. The MAP PSPL$+$GP model is also plotted for comparison. The data is detrended, meaning the MAP GP model realization (black curve in the second panel) is subtracted away. \emph{Fourth panel:} Residual flux after subtracting the PSPL$+$GP model. The blue curve shows the difference between PSPL$+$parallax$+$GP and PSPL$+$GP models. In this case the two are similar (PSPL$+$parallax$+$GP to PSPL$+$parallax$+$GP likelihood-ratio=1.6).}
    \label{fig:fig5}
\end{figure*}

In Figure \ref{fig:fig3} we presented the type of time-domain correlated noise that can cause troubles with fitting microlensing events with a model that has a constant baseline (such as the PSPL or PSPL$+$parallax models). In Figure \ref{fig:fig5} we show the MAP model (with the GP and PSPL$+$parallax components separated) along with samples from the posterior in the data-space for this same event. Though technically the microlensing parameters and GP kernel hyper-parameters are fit simultaneously, this plotting technique helps visualize how the model is working. The spread in the baseline flux for the random posterior samples is indicative of the combined uncertainty in the GP model and the baseline flux; however, the microlensing parameters are all well sampled with minimal auto-correlation. The PSPL$+$GP and PSPL$+$parallax$+$GP models are in good agreement for this event. 

\begin{figure*}
    \includegraphics[width=\textwidth]{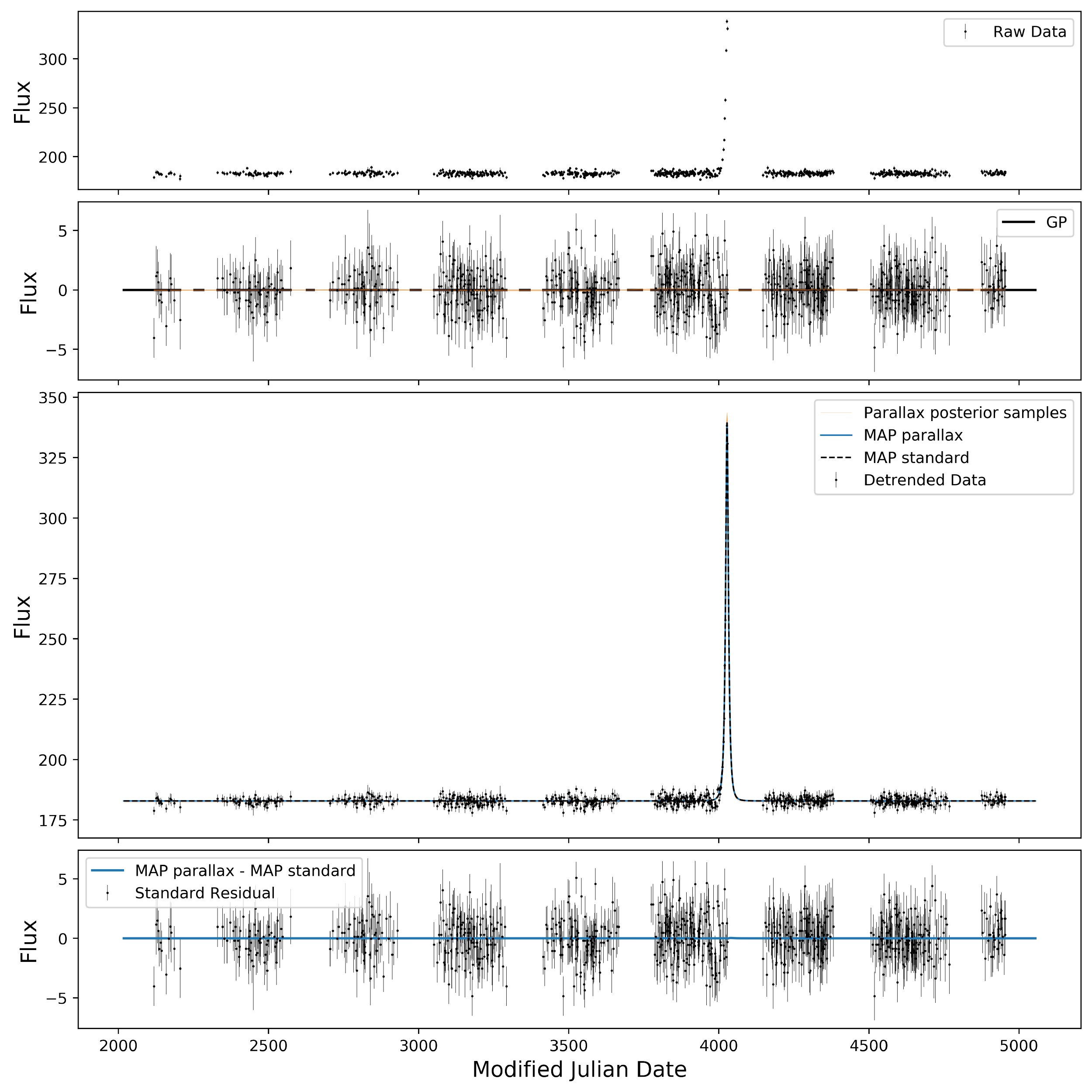}
    \caption{Example OGLE-III light curve (OGLE BLG 155.1.13052) with results from our MCMC analysis. This figure is analogous to Figure \ref{fig:fig5} but with a much simpler GP component. Again there is little parallax signal (PSPL$+$parallax$+$GP to PSPL$+$parallax$+$GP likelihood-ratio=1.1).}\label{fig:fig6}
\end{figure*}

In Figure \ref{fig:fig6} we show (analogously to Figure \ref{fig:fig5}) a light curve that has a much flatter baseline. Here the posterior samples are tightly clustered around the MAP in both the microlensing and GP panels, and the sampler has sufficiently mapped out the posterior for all parameters. This is due to the relatively high signal-to-noise ratio (SNR) of the data. Similar to the previous example, there is not a substantial difference between the PSPL$+$parallax$+$GP and PSPL$+$GP models. This is largely due to the short timescale of these events.

\begin{figure*}
    \includegraphics[width=\textwidth]{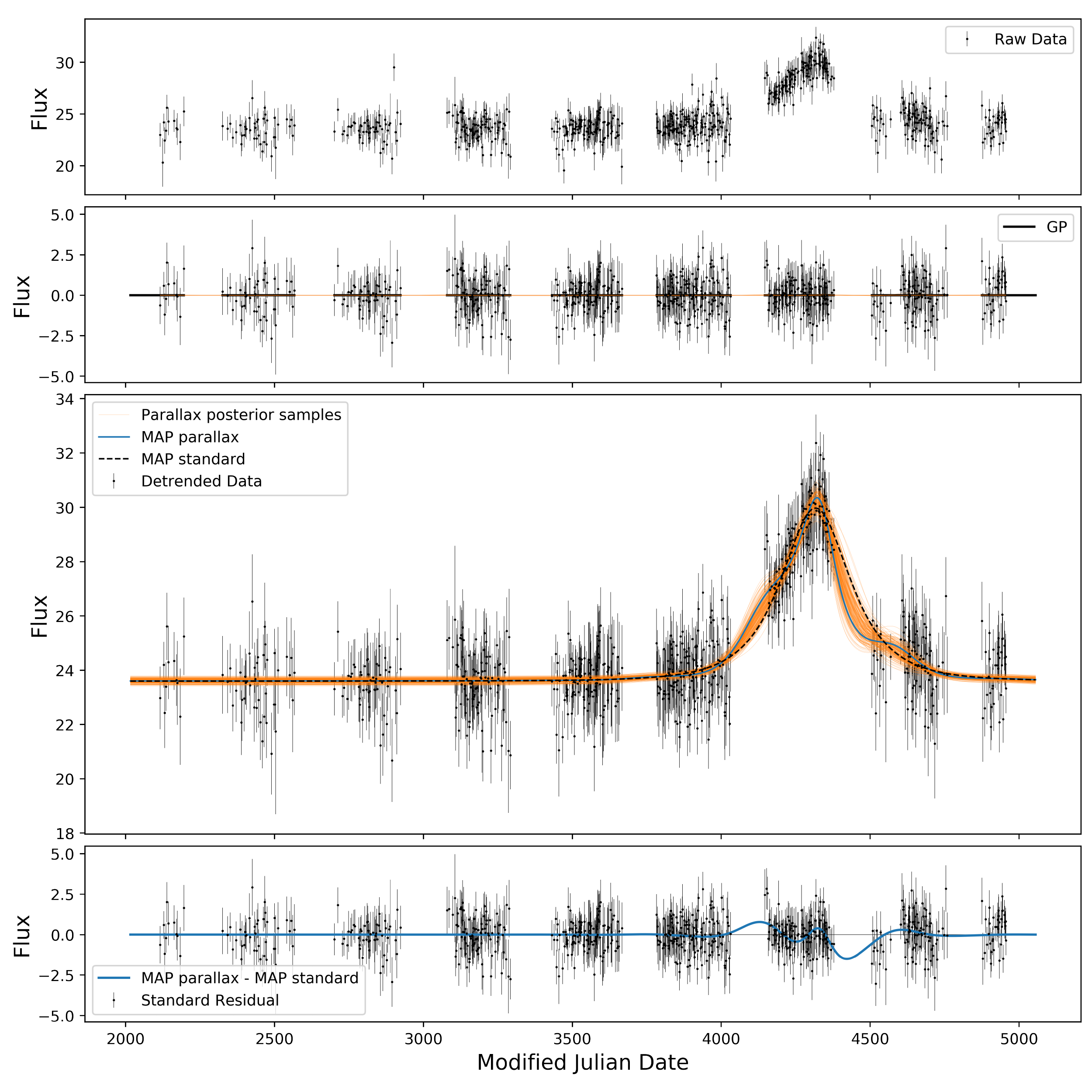}
    \caption{Example OGLE-III light curve (OGLE BLG 122.6.83113) with results from our MCMC analysis. This figure is analogous to Figure \ref{fig:fig5} but with a much simpler GP component and a clear parallax signal. Despite most of the parallax signal lying in the seasonal gaps in observability, the parallax model is still a better fit ((PSPL$+$parallax$+$GP to PSPL$+$parallax$+$GP likelihood-ratio=4.8).}\label{fig:fig7}
\end{figure*}

In Figure \ref{fig:fig7} we show (analogously to Figure \ref{fig:fig6}) another light curve that has a much flatter baseline flux; however, in this example there is clear parallax signal evident in the asymmetry of the microlensing event. This relatively long event with parallax signal demonstrates a limitation of observing Galactic bulge events, which are always un-observable for a portion of the year since the bulge is close to the Ecliptic and therefore periodically close to the Sun; the seasonal viewing means there are large gaps in the time series. This yearly pattern ensures larger uncertainty in the level of microlensing parallax since the residual from non-parallax models is also periodic on yearly timescales. Readers will notice the increased spread in the posterior samples (orange curves) in the third panel of Figure \ref{fig:fig7}. In \S\ref{subsec:BH_select} we discuss how this presents a challenge for selecting black hole microlensing events with photometric microlensing data. This event has a much lower SNR and has little signal in the GP model.

\begin{figure*}
    \includegraphics[width=\textwidth]{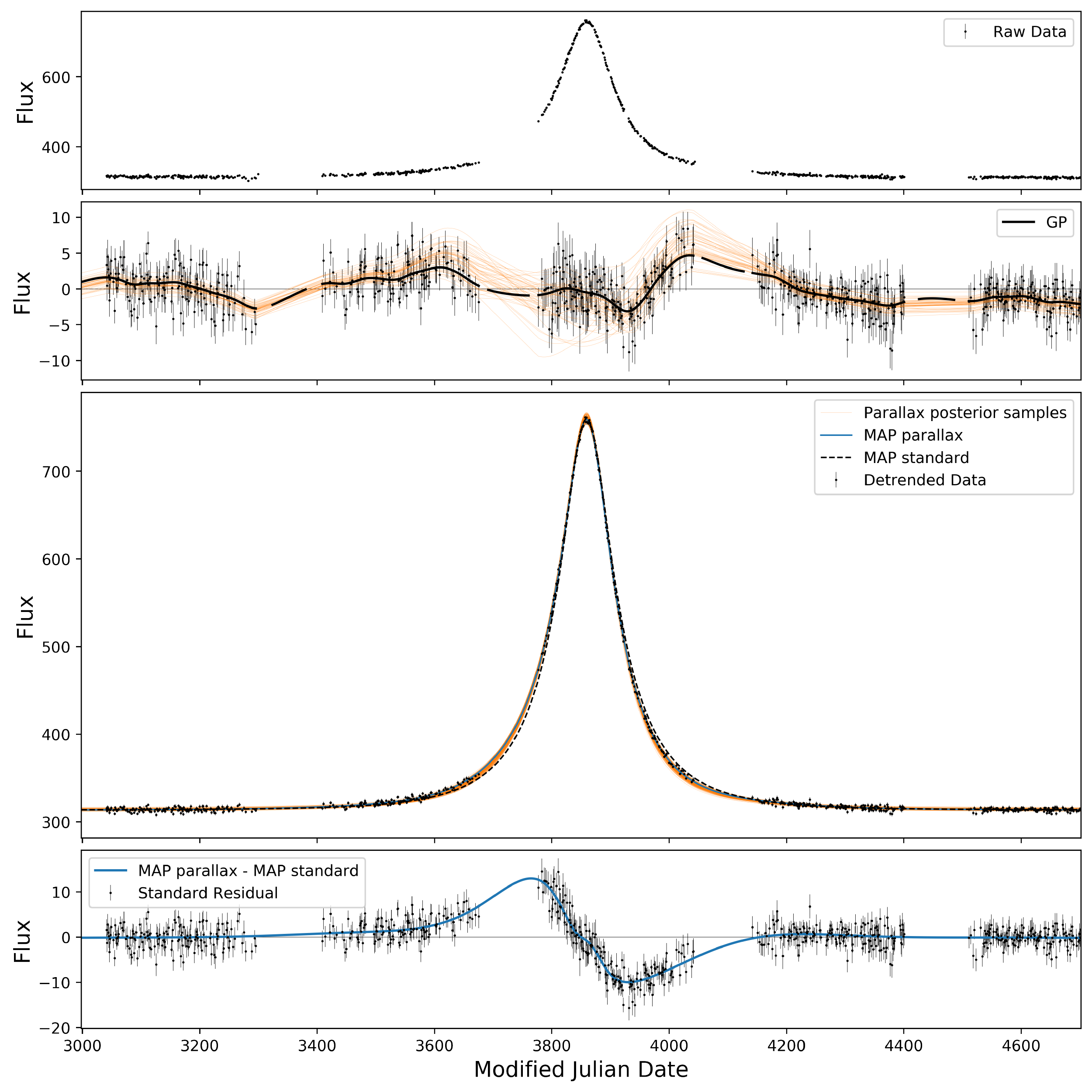}
    \caption{Example OGLE-III light curve (OGLE BLG 102.7.44461 with results from our MCMC analysis. This figure is analogous to Figure \ref{fig:fig5} but with a much simpler GP component and a clear parallax signal ((PSPL$+$parallax$+$GP to PSPL$+$parallax$+$GP likelihood-ratio=25).}\label{fig:fig8}
\end{figure*}

In Figure \ref{fig:fig8} we show (analogously to Figure \ref{fig:fig7}) a light curve that has both GP signal and parallax signal. Once again, the spread in the orange curves of the third panel are largest in the seasonal observing gaps. This event demonstrates the power of the simultaneous GP and microlensing inference. The second and fourth panels show that the flux of the stellar variability and microlensing parallax signals are similar in magnitude as well as coincident in time. Nevertheless, the full model is again well sampled and unaffected by auto-correlation. The parallax signal is clearly detectable over at least three years.

\begin{figure*}
    \includegraphics[width=\textwidth]{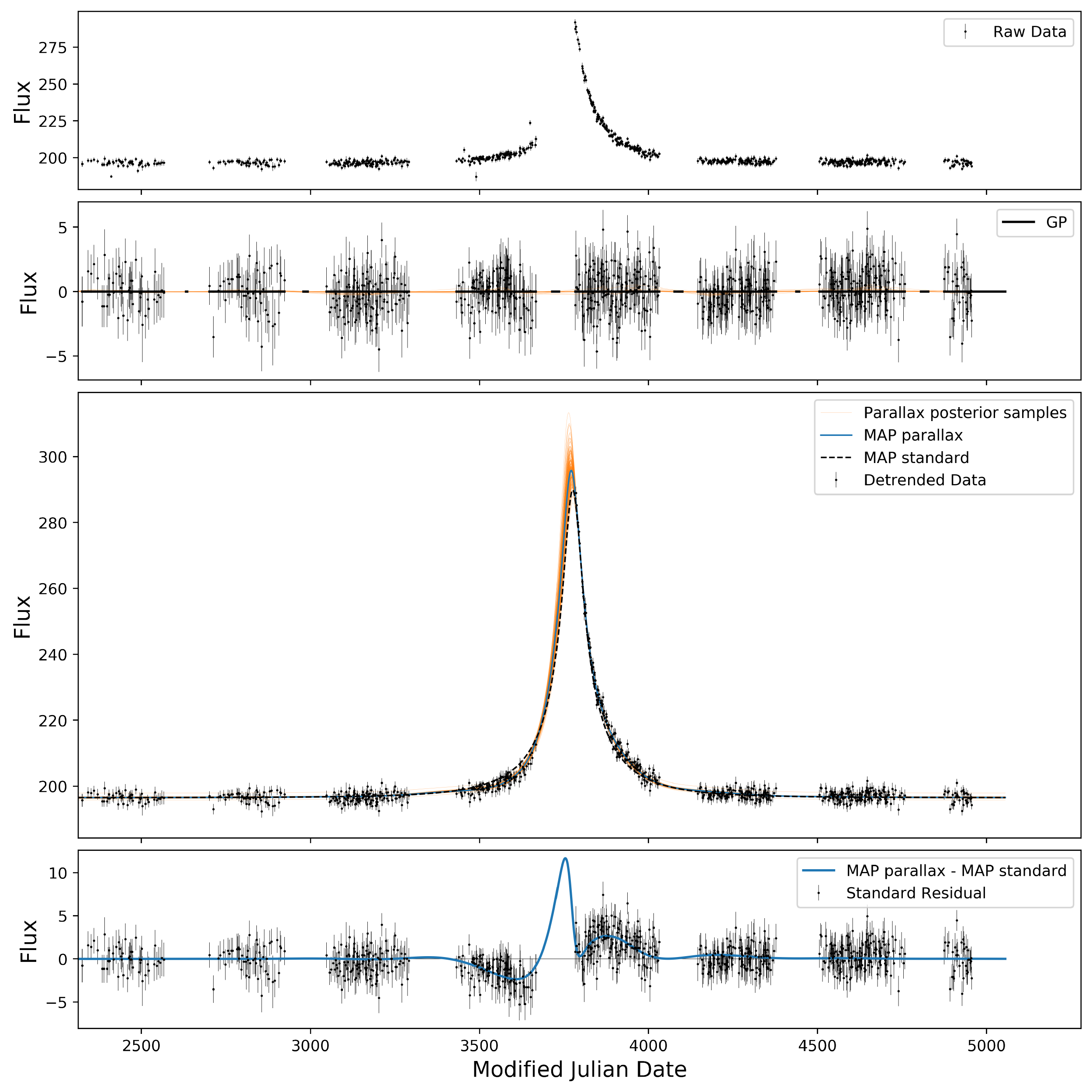}
    \caption{Example OGLE-III light curve (OGLE-III BLG 122.1.184151) with results from our MCMC analysis. This figure is analogous to Figure \ref{fig:fig5} and shows the results for the OGLE-III event with the largest Einstein crossing time. There is strong parallax (PSPL$+$parallax$+$GP to PSPL$+$parallax$+$GP likelihood-ratio=8.8), but again much is apparently missing as it occurs in a gap in the data.}\label{fig:fig9}
\end{figure*}

\begin{figure*}
    \includegraphics[width=\textwidth]{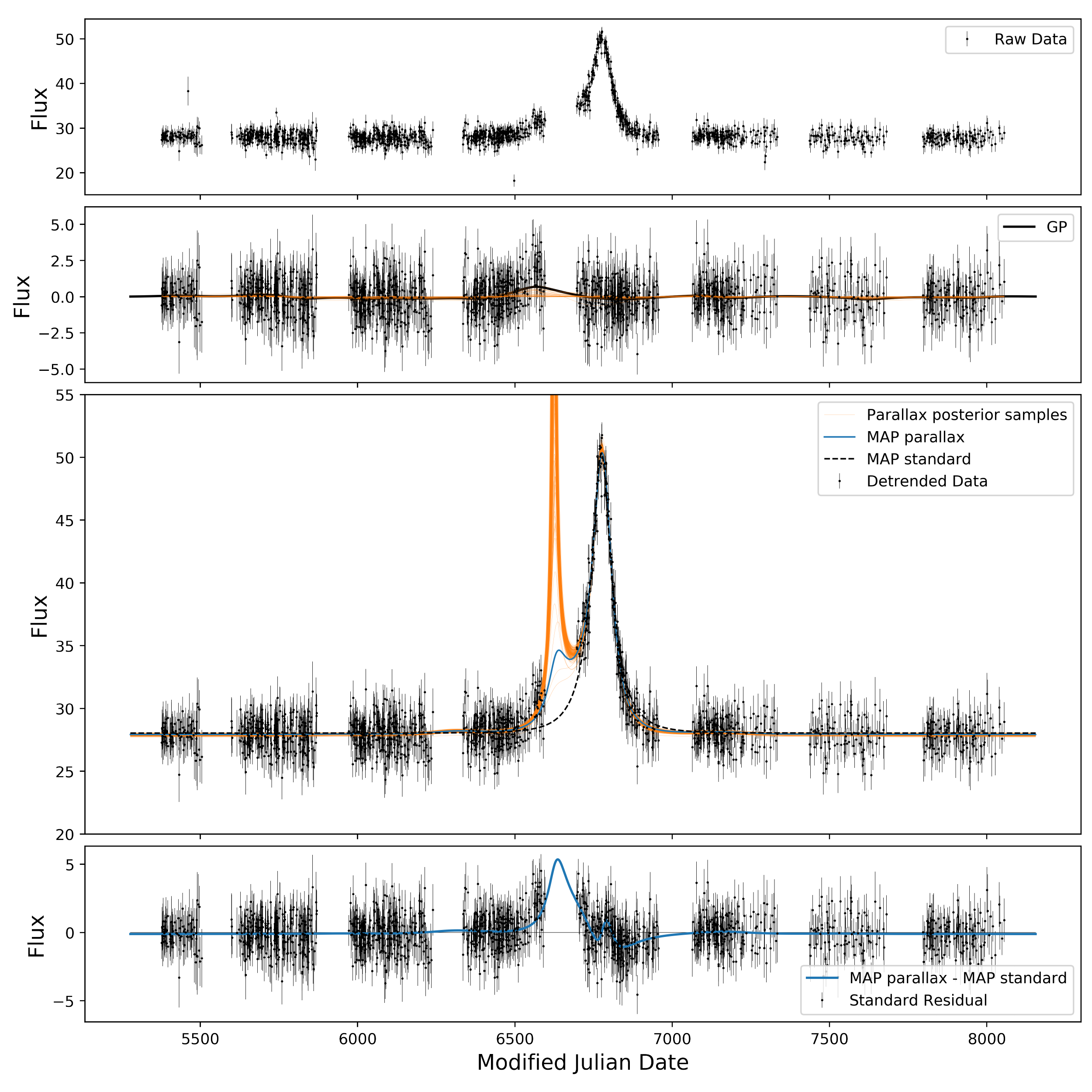}
    \caption{Example OGLE-IV light curve (OGLE-IV BLG 514.15.53029) with results from our MCMC analysis. This figure is analogous to Figure \ref{fig:fig5} except with OGLE-IV data from \citet{2019ApJS..244...29M}. It shows the results for the OGLE-IV event with the largest Einstein crossing time. This event has the most dramatic apparent parallax signal; however, the vast majority is fit within the observing gaps (PSPL$+$parallax$+$GP to PSPL$+$parallax$+$GP likelihood-ratio=3.0).}\label{fig:fig10}
\end{figure*}

In Figures \ref{fig:fig9} and \ref{fig:fig10} we present the longest timescale events in OGLE-III and IV from our modeling, respectively. For the OGLE-III event, most of the parallax signal is present in the seasonal observing gap; however, there remains parallax signal across consecutive observing seasons due to the long time scale. There is minimal baseline variability in this light curve. For the OGLE-IV event in Figure \ref{fig:fig10}, the majority of the posterior samples suggest a dramatic parallax signal with the highest peak occurring in the gap between observing seasons, once again demonstrating this challenging aspect of observing long duration microlensing events toward the Galactic bulge that exhibit strong parallax from a single site. This light curve also shows a discrepancy between the MAP and the majority of the posterior samples. There are only about 1\% of the posterior samples trending down toward the MAP realization which prefers a less dramatic peak in the data gap. Further inspection in the second panel of Figure \ref{fig:fig10} suggests that the GP signal in the MAP model is stealing a fraction of parallax signal. Again here, the vast majority of the posterior samples disagree with this picture. We discuss this in detail in \S\ref{subsec:gp_discuss}.

The samples for all twelve model parameters for 3560 OGLE-III and 5790 OGLE-IV are available in the online version of this paper.

\subsection{Population}\label{subsec:results_pop}
In \S\ref{sec:pop_model} we described our model the global $t_E$ distribution based on our samples for the individual microlensing event posterior PDFs. We chose a relatively simple model, a step function (i.e., histogram), over $\log_{10}t_E$. We fixed the number of bins to 30 log-spaced bins between 0.5 and 3000 days for linear $t_E$. We used a Dirichlet prior and sampled the relative bin heights for the OGLE-III and OGLE-IV populations separately. This is a necessary choice since the two surveys are not coincident on the sky and therefore will not necessarily have the same intrinsic $t_E$ distribution. 

\begin{figure*}
    \includegraphics[width=\textwidth]{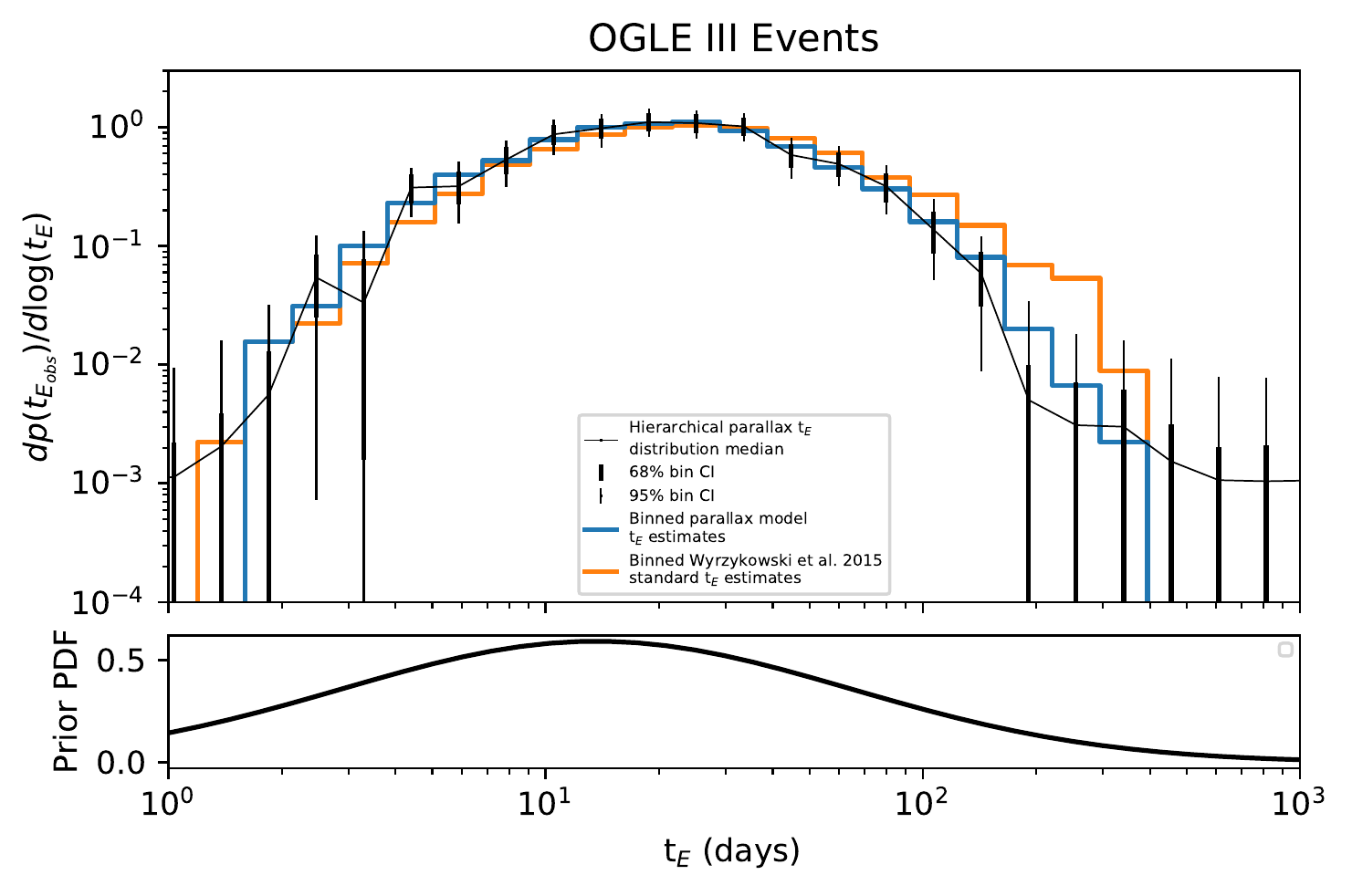}
    \caption{OGLE-III population results for the global observed $t_E$ distribution (no survey efficiency corrections have been applied to any of the distributions). The black thin and thick bars show the $68\%$ and $95\%$ confidence intervals on the bin heights in our hierarchical model of the global distribution. The orange histogram shows the binned values presented in \citet{2015ApJS..216...12W} for comparison, and the blue histogram shows the MAP point estimates from our individual PSPL$+$parallax$+$GP model fits. In the lower panel, we show the prior PDF on $t_E$ given by Eq.\ \ref{eq:tE}. Note that the effect of our broadening of the $t_E$ prior is apparent by the amplitude that varies from $\sim0.1--0.5$, whereas the actual distribution amplitudes varies by approximately 3 dex.} \label{fig:fig11}
\end{figure*}

\begin{figure*}
    \includegraphics[width=\textwidth]{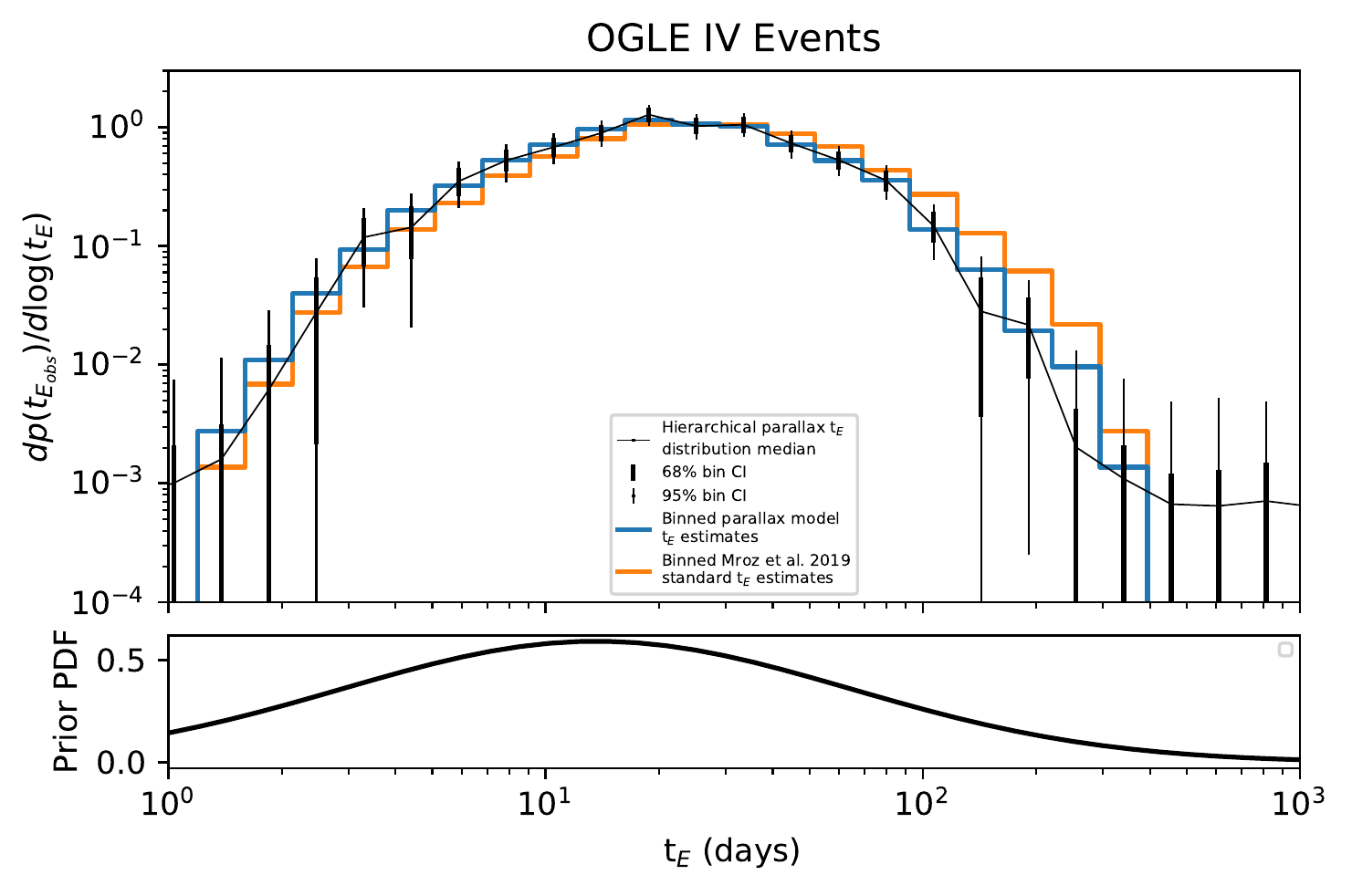}
    \caption{OGLE-IV population results for the global observed $t_E$ distribution (no survey efficiency corrections have been applied to any of the distributions). The black thin and thick bars show the $68\%$ and $95\%$ confidence intervals on the bin heights in our hierarchical model of the global distribution. The orange histogram shows the binned values presented in \citet{2019ApJS..244...29M} for comparison, and the blue histogram shows the MAP point estimates from our individual PSPL$+$parallax$+$GP model fits. In the lower panel, we show the prior PDF on $t_E$ given by Eq.\ \ref{eq:tE}. Note that the effect of our broadening of the $t_E$ prior is apparent by the amplitude that varies from $\sim0.1--0.5$, whereas the actual distribution amplitudes varies by approximately 3 dex.}\label{fig:fig12}
\end{figure*}

The results for our OGLE-III and OGLE-IV population $t_E$ distribution modeling are presented in Figures \ref{fig:fig11} and \ref{fig:fig12}, respectively. For OGLE-III, the results are plotted as a series of confidence intervals on the bin heights in comparison to the histogram of $t_E$ values published by \citet{2015ApJS..216...12W} for their PSPL model fits to the OGLE-III events (orange bins). We also plot the histogram of MAP $t_E$ values for our PSPL$+$parallax$+$GP model fits (blue histogram). The bins are in general agreement for short time-scale events but start to diverge at greater than $2\sigma$ around 100 days, with PSPL fits showing more events with longer time-scales. This remains true for all bins corresponding to $t_E>100$ days. The hierarchical modeling is in general agreement with the MAP $t_E$ histogram for our PSPL$+$parallax$+$GP fits across the domain.

For OGLE-IV (Figure \ref{fig:fig12}), the results are plotted similarly except the orange histogram corresponds to the distribution of PSPL model $t_E$ point estimates from \citet{2019ApJS..244...29M}. The results for our OGLE-IV analysis largely follow the same trends as the OGLE-III results. We find evidence for fewer long time-scale events as compared to the histogram of point estimates from the previous OGLE analysis, which again assumed a simple PSPL model.


\section{Discussion}\label{sec:discussion}
In this section we have categorized some of the main points that were mentioned as worthy of further discussion throughout the text above as well as a handful of key concerns the astute reader may have raised thus far. 

\subsection{Gaussian Process Model Fitting}\label{subsec:gp_discuss}
\begin{figure*}
    \includegraphics[width=\textwidth]{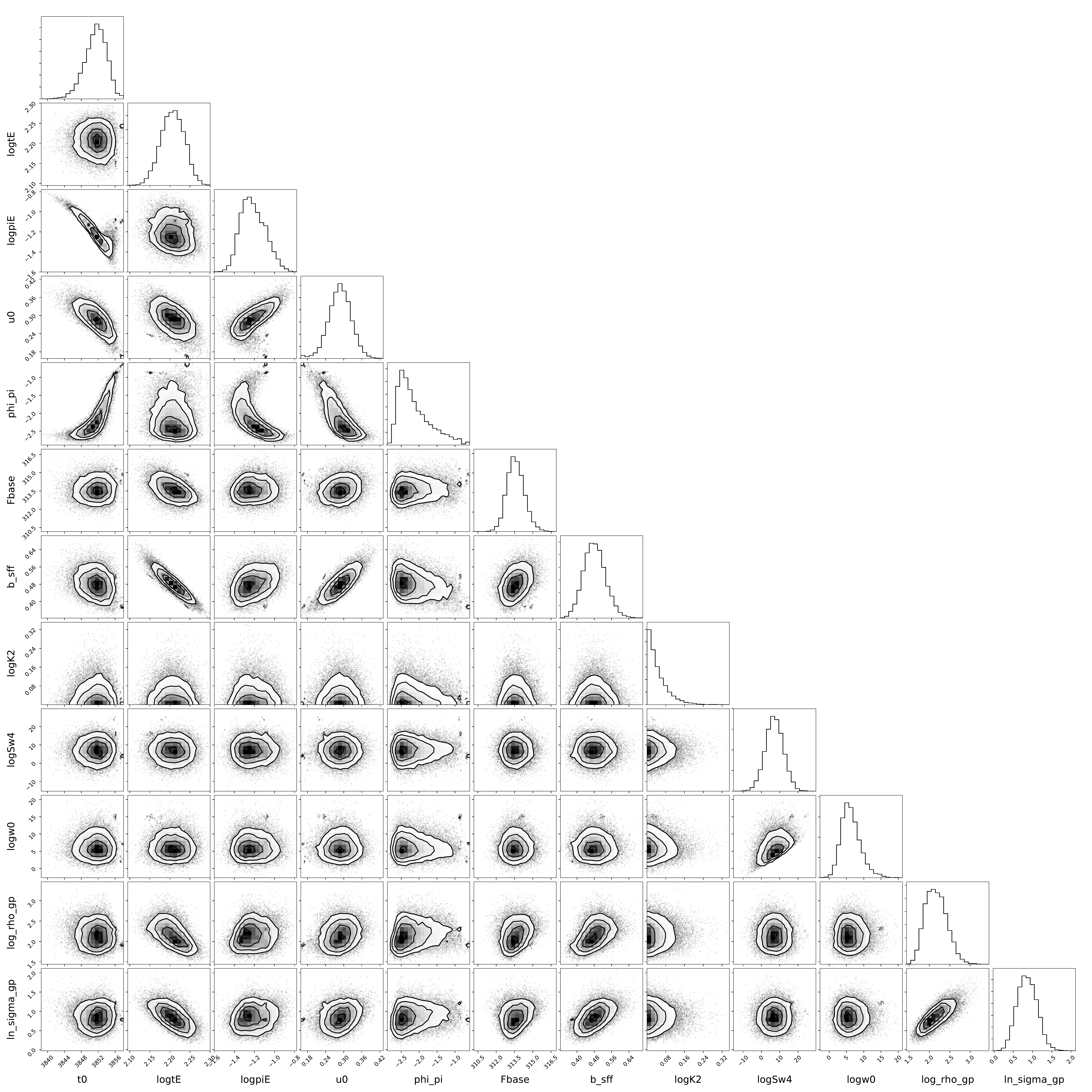}
    \caption{Posterior corner plot for OGLE-III BLG 102.7.44461. The strongest correlations are between microlensing parameters. There is a slight correlation between $t_E$ and the the Mat\'{e}rn-3/2 kernel parameters; however, in general, the GP hyper-parameters are much more well-behaved than the microlensing parameters.}\label{fig:fig13}
\end{figure*}

The model parameters we have sampled include both the microlensing parallax and GP hyper-parameters. These were fit simultaneously, meaning we can estimate correlations between them. This is important because one of the key worries of utilizing GP models in microlensing is the concern that the GP model might absorb some or all of the actual microlensing signal. This is of further concern for our model since the yearly parallax signal causes a quasi-periodic deviation from the PSPL model curve, and our GP was designed to handle quasi-periodic signals in the baseline. In Figure \ref{fig:fig8} we showed an example fit with temporally coincident GP and parallax signal of approximately the same magnitude. In Figure \ref{fig:fig13}, we show the posterior corner plot for this event, which shows the full set of one and two dimensional marginal posterior PDFs for all samples in the chain. By and large, the GP hyper-parameters are more well behaved than the microlensing parameters; however, there is a slight correlation between $t_E$ and the Mat\'{e}rn 3/2 parameters. This despite no obvious acceptance of parallax signal by the GP model in Figure \ref{fig:fig8}.

In order to explore the effect of including the GP model in our likelihood function, we injected and fit microlensing events. We selected the 50 OGLE-IV events with the smallest median $t_E$ value. We masked the data between $t_0\pm5 t_E$ and injected microlensing events drawn randomly from our prior distributions onto the masked baseline. We re-scaled the uncertainty in the remaining flux measurements assuming Poisson noise. We preserved the . We fit these synthetic data with our model and computed the relative bias as:
\begin{equation}\label{eq:bias}
    \text{bias} = \frac{\boldsymbol{\theta}_{\text{med}}-\boldsymbol{\theta}_{\text{inj}}}{\sigma_{\boldsymbol{\theta}}}
\end{equation}where the difference in the numerator is between the marginal inferred median value and the injected truth for each parameter, and the denominator indicates the standard deviation in the chains for each parameter.

We present the results from this analysis in Table \ref{tab:injection} as the parameters of a Gaussian fit to the bias values for each of the microlensing parameters. If there is no inherent bias, we expect the bias values to be normally distributed with zero mean and unit variance. We find that for most of the microlensing parameters this holds approximately true. 

\begin{table}
\centering
\caption{Summary of injection analysis}\label{tab:injection}
\begin{tabular}{lll}
Parameter & $\mu_\mathrm{bias}$ & $\sigma_\mathrm{bias}$\\
\hline
$F_{\mathrm{base}}$ &   $-6.8\times10^{-4}$ & 0.040 \\
$b_{\mathrm{sff}} $ &   0.32                & 1.1 \\
$t_{0}$             &   $-7.8\times10^{-3}$ & 1.4 \\
$\log_{10} u_{0}$   &   0.16                & 1.3 \\
$\log_{10} t_E$     &   -0.12               & 0.52 \\
$\log_{10} \pi_E$   &   0.11                & 0.77 \\
$\phi_{\pi}$        &   -0.029              & 0.97 
\end{tabular}
\tablecomments{Here $\mu$ and $\sigma$ refer to the mean and standard deviation of the bias (see Eq. \ref{eq:bias}) for each parameter listed.}
\end{table}

There is a slight positive bias for $b_{\text{ssf}}$ and the spread in relative bias for $F_\text{base}$ is much smaller than expected. The relative bias in $b_\text{sff}$ is largely due to poor inference in highly blended events. That is, when the injected value of $b_{\text{sff}}$ is close to zero, the posterior tends to have a large spread, and the median value tends toward the median of the prior, which is closer to 0.5.  On the other hand, the small spread in the bias for $F_\text{base}$ is due to an enhancement in the denominator of Eq. \ref{eq:bias} for that parameter due to the GP model. An example of this is present in Figure \ref{fig:fig5}, where the uncertainty in the GP model manifests additional uncertainty in the baseline flux. The positive bias in $u_0$ may also be of concern, but this is again due to a one-sided prior (bounded at zero), and a tendency for low SNR events to trend toward the median of the prior. Note that if we had included negative $u_0$ models, there would be a dichotomy in this bias (negative for $u_0<0$ injected models and positive for $u_0>0$. We again note that we checked for a bias in our $t_E$ population modeling by including only $u_0>0$ and found none. This follows from the geometry.

\begin{figure*}
    \includegraphics[width=\textwidth]{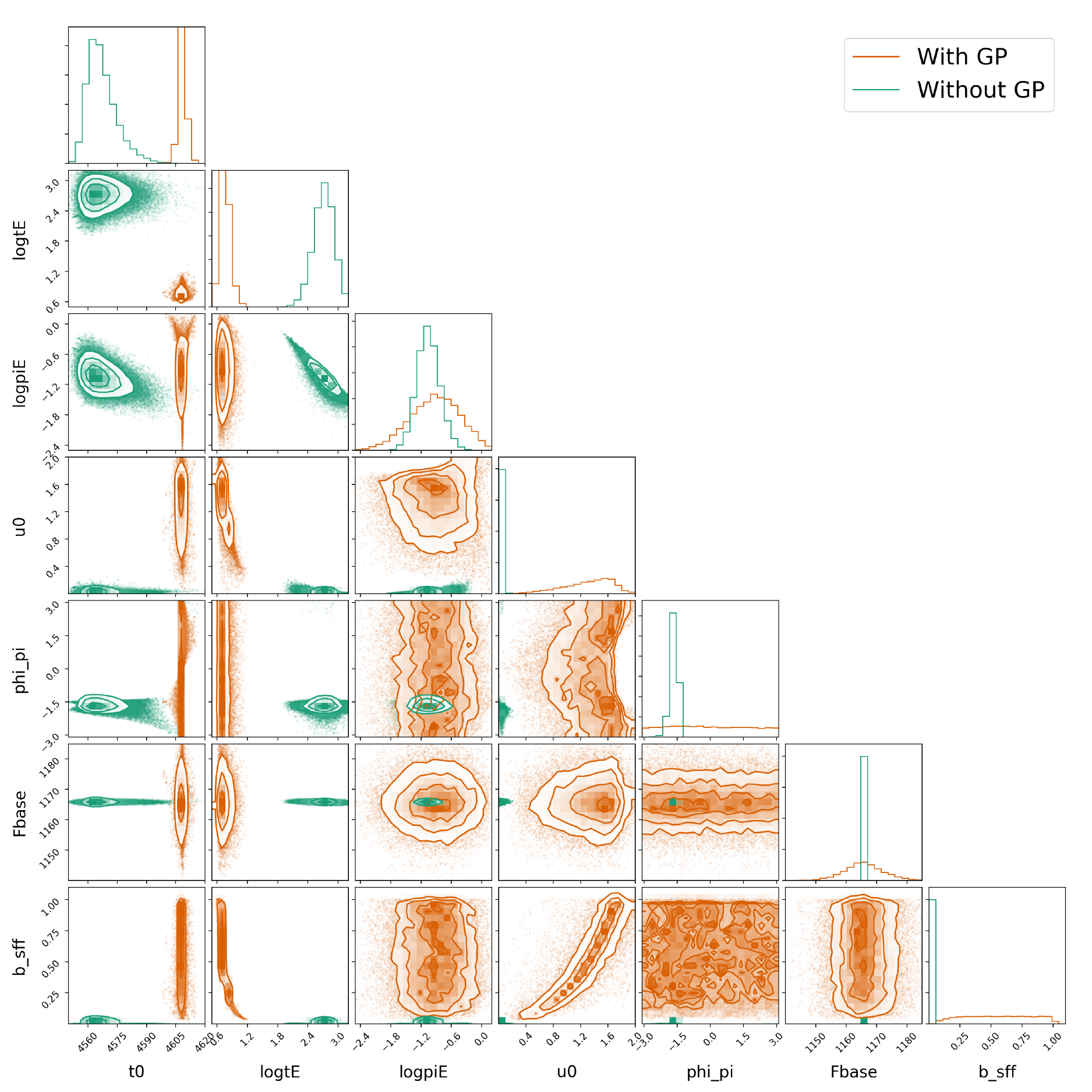}
    \caption{Posterior corner plot for OGLE-III BLG 156.7.141434 (see Figure \ref{fig:fig5}) covering the microlensing model parameters with and without the GP component of the model.}\label{fig:fig16}
\end{figure*}

\begin{figure*}
    \includegraphics[width=\textwidth]{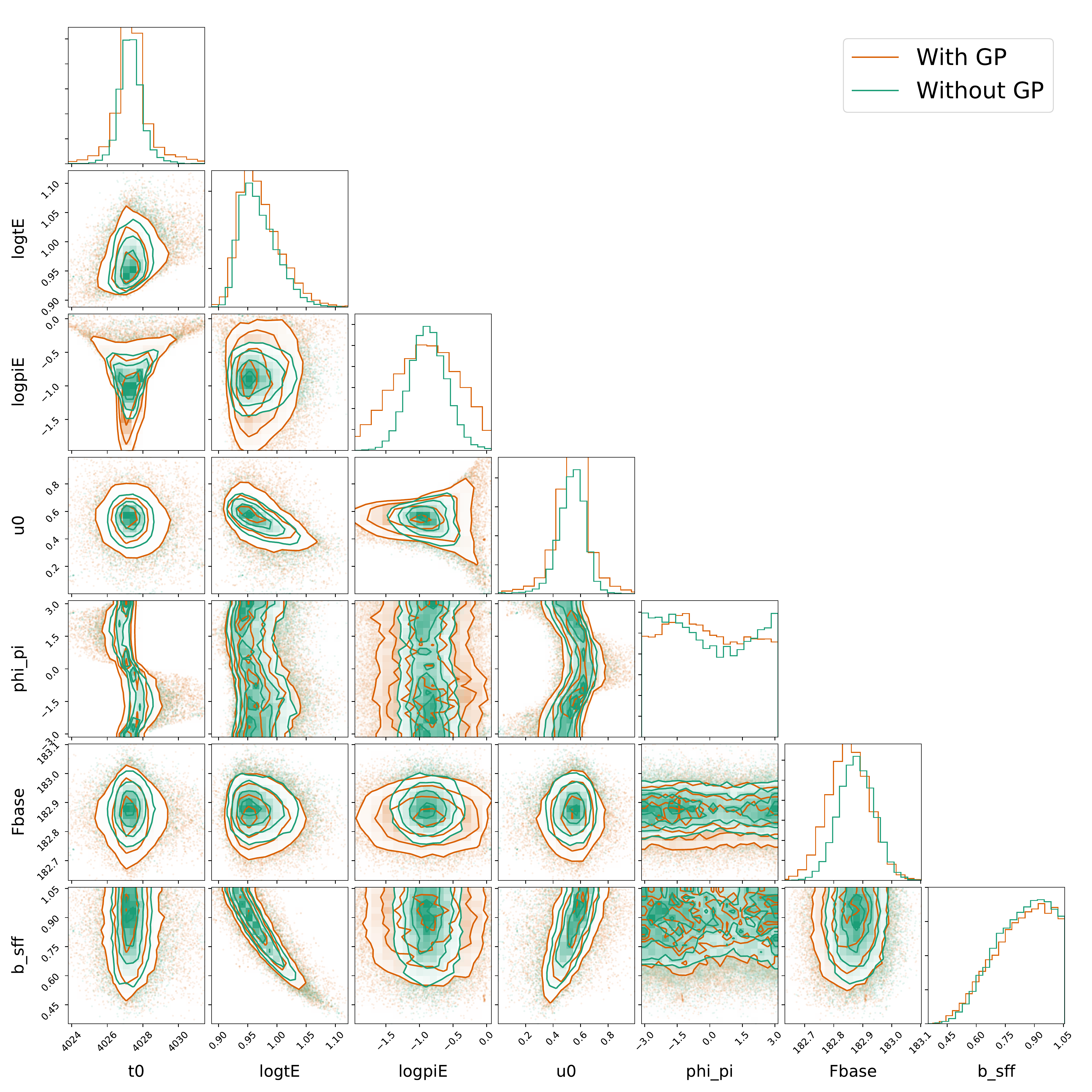}
    \caption{Posterior corner plot for OGLE-III BLG 155.1.13052 (see Figure \ref{fig:fig6}) covering the microlensing model parameters with and without the GP component of the model.}\label{fig:fig14}
\end{figure*}

\begin{figure*}
    \includegraphics[width=\textwidth]{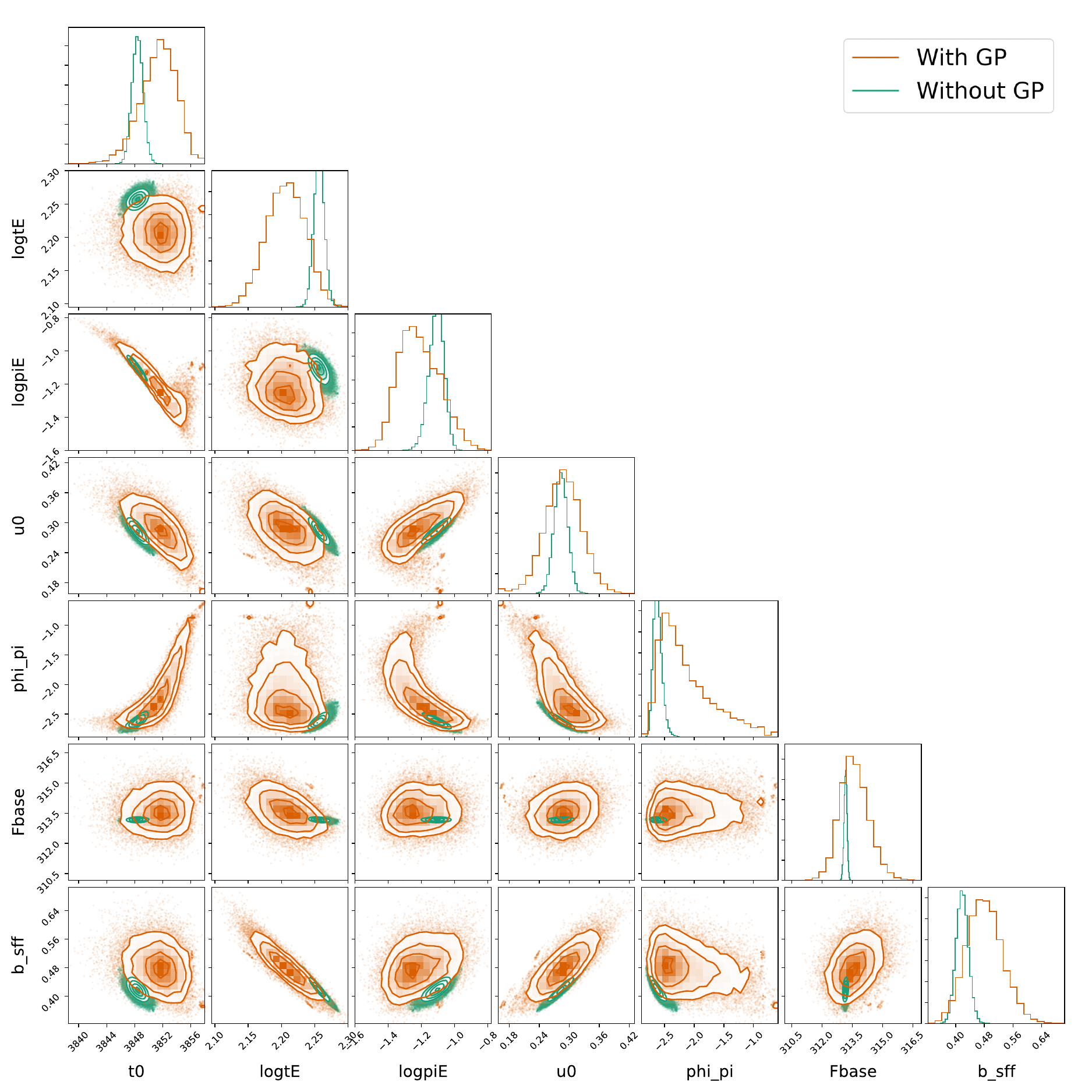}
    \caption{Posterior corner plot for OGLE-III BLG 102.7.44461) (see Figures \ref{fig:fig8} and \ref{fig:fig13}) covering the microlensing model parameters with and without the GP component of the model.}\label{fig:fig15}
\end{figure*}

Another test for our GP inference is to remove it and fit the same data with only the microlensing model. For the majority of our events there is good agreement, since the baseline flux is generally smooth; although, in some cases, there are catastrophic breakdowns, which was the impetus for developing this GP in the first place. We did this for the OGLE-III events shown in Figures \ref{fig:fig5}, \ref{fig:fig6} and \ref{fig:fig8} to demonstrate a range of scenarios. In Figure \ref{fig:fig14} we show overlapping posteriors for models with (orange) and without (green) the GP component of our model for the light curve shown in Figure \ref{fig:fig5}. This demonstrates the importance of including a GP for events with strong variability in the baseline flux. Without the GP model, the microlensing model is forced into a corner of parameter space that has nearly zero flux from the source, a near-zero impact parameter and an extremely long timescale despite this clearly not being the case upon visual inspection. It is also clear that the modeling including the GP model is handling parallax better. Given the short timescale of the event, there is little ability to estimate $\pi_E$ or $\phi$, so the sampler explored the entire prior distribution. In Figure \ref{fig:fig15}, which corresponds to the light curve in Figure \ref{fig:fig6}, we show that when there is no baseline variability, the GP and non-GP posteriors are in excellent agreement. As a final example, we present Figure \ref{fig:fig16} corresponding to the light curve shown in Figure \ref{fig:fig8}, which has both baseline variability (albeit much less than the case shown in Figures \ref{fig:fig5}) and parallax signal. Here we found the posterior PDFs for all parameters were inflated with the inclusion of the GP into the model. This may seem counter-intuitive since one might expect the more complicated GP model to be able to estimate physical parameters more tightly, but it is expected since the GP model propagates additional uncertainty into the model. Similar to Figure \ref{fig:fig14}, there is significant bias in the no-GP model, however to a lesser degree, suggesting that the bias is correlated with the level of intrinsic stellar variability. 

Figures \ref{fig:fig14}, \ref{fig:fig15} and \ref{fig:fig16} show the posterior PDFs for the two models for the microlensing parameters in common. There is an additional parameter we left out, $\log\left(K^2\right)$, since it was vastly different for the two models in light curves with baseline variability. In the absence of the GP model, this parameter is inflated in an attempt to mask over the variable baseline. The additional modeling flexibility of the GP is `more correct' in turning model biases into additional propagated uncertainty (the bias/variance trade-off). Also evident is the inflated $F_\text{base}$ parameter in the GP model, especially in Figure \ref{fig:fig5}. Cases such as this caused the low $\sigma$ value for that parameter in Table \ref{tab:injection}. Figures \ref{fig:fig14} and \ref{fig:fig16} show the value of adding a GP to model the variable baseline. Without it, the model is overly confident and often biased (as is the case here-- note the green posteriors in Figure \ref{fig:fig16} are often at the edge of the larger orange posteriors). In more catastrophic scenarios, the model is forced into extremely unlikely scenarios (as in Figure \ref{fig:fig14}. Only when there is an intrinsically flat baseline, as in the best case scenario (Figure \ref{fig:fig15}), does the GP inference agree with the non-GP modeling.

\subsection{Multi-modality of individual posterior PDFs}
It is well documented that the parallax signal in photometric microlensing data leads to a multi-modal likelihood \citep{2002MNRAS.336..670S,2004ApJ...606..319G,2005MNRAS.361..128S,2005ApJ...633..914P}. There are numerous degeneracies that have different values for microlensing parameters. These issues mostly arise for short time-scale events. It is typical in the literature to identify these degeneracies and initialize sampling algorithms in these sub-peaks and present the maximum likelihood models for each of the degenerate solutions. In our analysis, we have not gone through this procedure since we are not interested in the peculiarities of any single event. 

The degeneracies emerge from ambiguities on the components of the relative proper motion vector parallel and perpendicular to the Earth's acceleration. Since this is purely geometric and randomly distributed over the thousands of OGLE-III and OGLE-IV events, we allow our sampler to settle in a random mode for individual events by initializing the chains randomly under the prior PDF. The differences in the modes of any single event do not affect the population analysis, especially for long duration events ($t_E>100$ days) where the difference in $t_E$ for different modes is fractionally small, especially for low $\pi_E$ events which are most likely to be black holes (see \S\ref{subsec:BH_select}). However, we do caution against downloading the chains for an individual event and making inference on that event without a fuller exploration of the posterior. 

\subsection{The $t_E$ distribution}

The peaks of the $t_E$ distribution we estimated using the samples of all our of fits largely agree with the histograms of point estimates presented by \citet{2015ApJS..216...12W,2019ApJS..244...29M} -- this despite our prior having a lower peak timescale. Our prior was constructed from the OGLE-like survey simulated by \citet{2020ApJ...889...31L} where it was found that two different Galactic bar angles could match either the observed stellar density map or the microlensing event rates; but not both. In all cases simulated, the $t_E$ distribution peaked at shorter timescales than observed distributions (see also Medford et al. submitted). As can be seen from the bottom panels of Figures \ref{fig:fig11} \& \ref{fig:fig12}, we significantly broadened the \textsf{PopSyCLE} based prior to reduce the impact of this discrepancy.

There are significantly fewer long time-scale events across a range of timescales above 100 days in our $t_E$ distribution as compared to both the OGLE-III and IV results. However, both distributions also suggest there are multiple events with $t_E>400$ days despite none reported in either point-estimate histogram. This is due to the accumulation of high $t_E$ tails adding weight to these bins. Note that both \citet{2015ApJS..216...12W,2019ApJS..244...29M} apply a cap on the maximum PSPL $t_E$ point estimates of a few hundred days during the event detection and distillation process. We recommend not applying such long $t_E$ selection limits in the future, as they greatly impact the ability to conduct black hole studies.

\begin{figure*}
    \includegraphics[width=\textwidth]{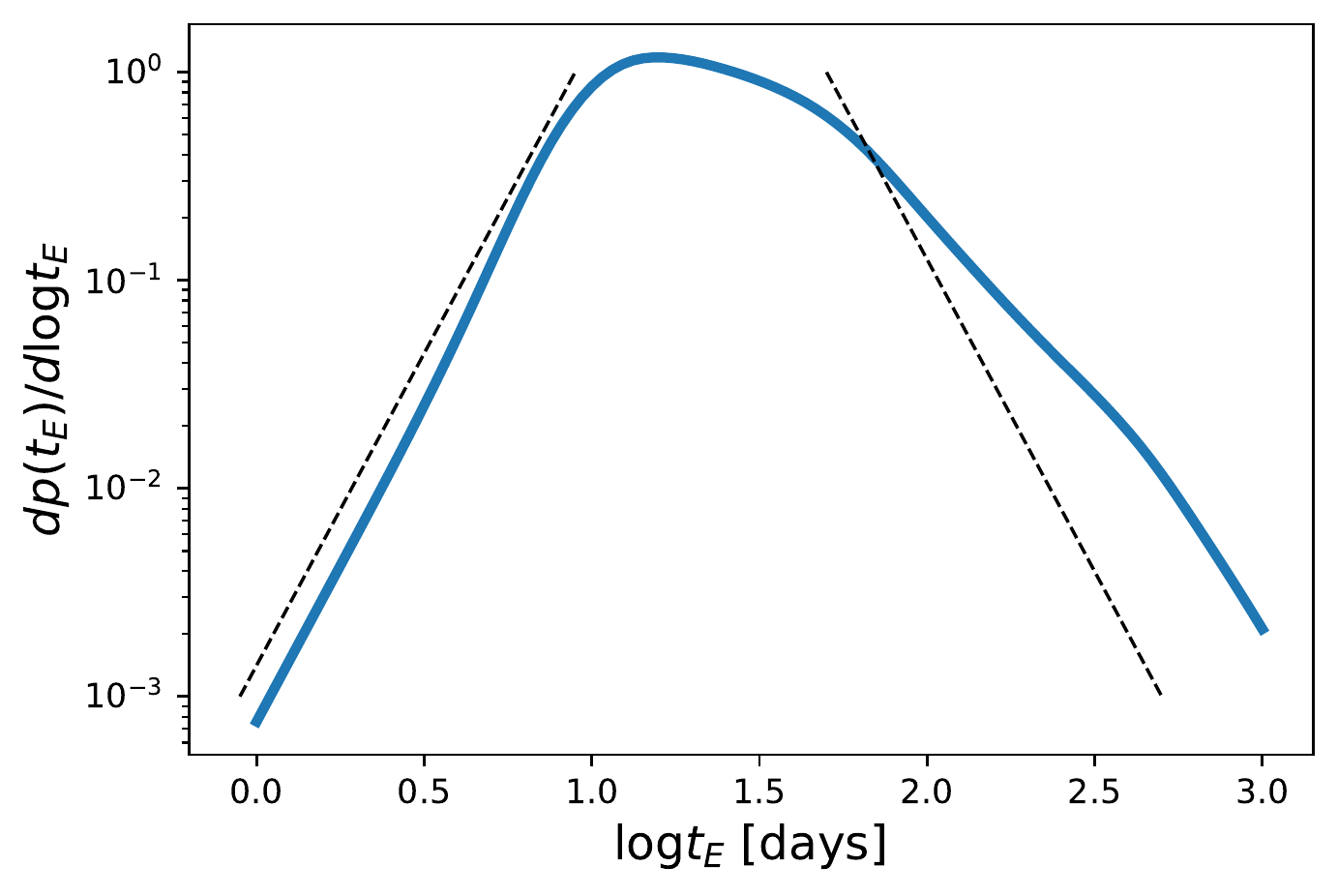}
    \caption{An example $t_E$ distribution from the \citet{1996ApJ...473...57M} formalism modified for a mixture model composed of two separate power-law mass distributions (blue curve). Mixture component 1 has a min and max mass range of $0.1-3\ M_\odot$ and power-law slope of -1.5, while mixture component 2 has a min and max mass range of $1-100\ M_\odot$ and power-law slope of -1.5. The relative proportion of each mixture is 0.8 and 0.2, respectively. The black dashed lines plot $\pm3$ power-law slopes. There is clearly deviation from the -3 slope due to the second mass component.}\label{fig:fig17}
\end{figure*}

Previous studies of the $t_E$ distribution often make reference to the slope of the short and long timescale rise of the distribution and the position of the peak. They especially note agreement with theory \citep[e.g.,][]{1996ApJ...473...57M} if the slopes of the $\log t_E$ distribution tails are $\pm3$ \citep[e.g.,][]{2015ApJS..216...12W,2019ApJS..244...29M}. This ``expectation'' for  $\log t_E=\pm3$ slopes is an oversimplification of the potential \citet{1996ApJ...473...57M} model diversity. While \citet{1996ApJ...473...57M} show that this is true for many choices of present-day mass function and galactic velocity distributions. They also show that it is not a universal tendency (see e.g., Figures 1 and 3 of their paper). The choice of slope ($\alpha$) for the present-day mass function can greatly affect both the high and low-end $t_E$ slope, with the potential for the effect to increase with wider mass distributions. Note though that, for single power-law mass models, varying $\alpha$ usually only causes $t_E$ slope deviations from $\pm3$ on one end or the other. For example, a power-law mass distribution with a steeper negative slope (e.g., $\alpha=-2.5$; i.e., more total mass in low mass objects) will maintain the +3 slope on the low-$t_E$-end, however will result in a shallower negative slope on the high-$t_E$-end (i.e., slope $>-3$). Notably, their model with $\alpha=-2.5$ $\beta = 2$ shifted the peak of the $t_E$ distribution to the left, which is evident in models presented by \citet{2020ApJ...889...31L}. More complex mass distributions, such a bi-modal distributions, can also cause deviations in the slopes \citep[see Figure \ref{fig:fig17} and][]{2019RNAAS...3...58L}.

Based on \citet{1996ApJ...473...57M} and our arguments above, it is not clear how it is physically possible to produce a slope steeper than $\pm3$\footnote{
\citet{2015ApJS..216...12W,2019ApJS..244...29M} only provide $t_E$ efficiency corrections based on point-source point-lens models without parallax, means we are not able to reliably apply their efficiency corrections to our parallax based $t_E$ distribution (additionally, as \citet{2019ApJS..244...29M} note, their corrected distribution will be biased at the high $t_E$ end). Furthermore, without access to the full OGLE datasets, we are unable to develop appropriate efficiency corrections. Thus, care should be taken when interpreting our $t_E$ distributions in terms of the \citet{1996ApJ...473...57M} formalism. That said, both \citet{2015ApJS..216...12W,2019ApJS..244...29M} suggest rather flat efficiency corrections at the high-$t_E$ end so it seems reasonable to consider the high-$t_E$ slope with respect to the theoretical expectations for the slope.
}.
Note that the high-$t_E$ slopes of our distributions are consistent with $-3$ although favoring slightly steeper slopes. This may be indicative of bias due to the long $t_E$ selection cuts of \citet{2015ApJS..216...12W,2019ApJS..244...29M} applied to the $t_E$ \citet{1986ApJ...304....1P} model parameter.
\citet{1996ApJ...473...57M} also note that the long tails of the $t_E$ distribution will also be modified by the motion of the Earth.
However, when we use the \citet{1986ApJ...304....1P} model with GP we find a distribution more consistent with our parallax+GP model distribution, suggesting that the bias may primarily be due to not modeling the intrinsic stellar variability, as we discussed in \S\ref{subsec:gp_discuss}.

\subsection{Selecting Black Hole Candidates}\label{subsec:BH_select}
Black hole microlensing has not been mainstream science since the early constraints on MACHO dark matter in the 1990s and 2000s. There have been searches for black holes in the MACHO Survey \citep{2002ApJ...579..639B,2002ApJ...576L.131A,2005ApJ...633..914P} and OGLE-III \citep{2016ApJ...830...41L,2016MNRAS.458.3012W,2020AA...636A..20W} since then, but as MACHOs went out of style as a dark matter candidate, focus instead was placed on detecting astrophysical black holes. Interest further heightened for both types of black holes with LIGO as it revealed both higher-than-expected black hole masses \citep{2016PhRvL.116t1301B} without resolving the possible mass gap between the neutron star and black hole populations \citep{2015ApJ...807L..24L}.

\begin{figure*}
    \includegraphics[width=\textwidth]{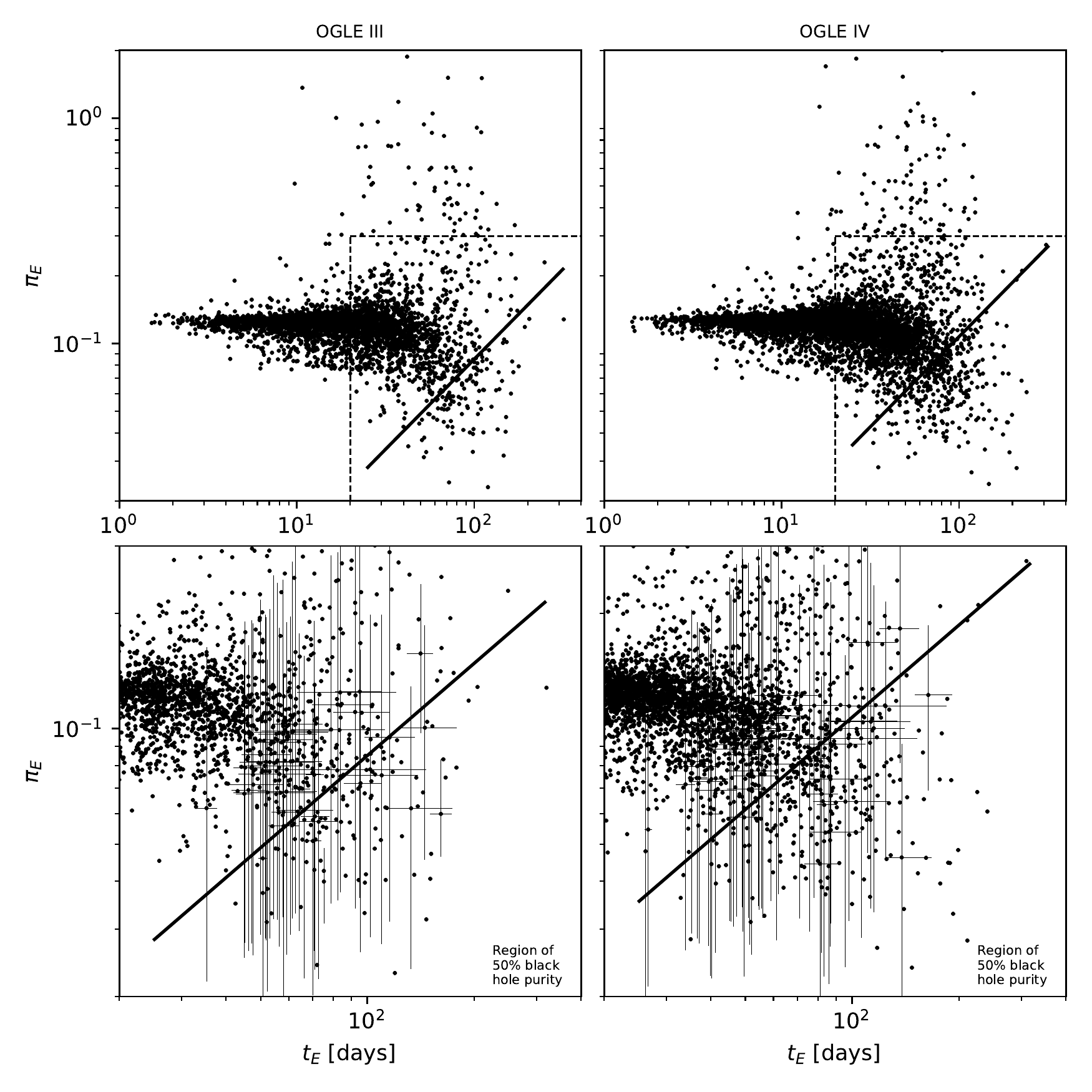}
    \caption{\emph{Top row:} Marginal $t_E$--$\pi_E$ values for 3560 OGLE-III (left panel) and 5760 OGLE-IV (right panel) microlensing events. The points represent the median chain values for the two parameters. The lines demarcate 50\% completeness of astrophysical black holes in the \textsf{PopSyCLE} OGLE-III and IV simulations \citep{2020ApJ...889...31L}. There are 107 and 283 events below the lines for OGLE-III and IV, respectively. The dashed box is presented as an inset below. \emph{Bottom row}: Zoom into the black hole demarcation line. We've presented the same data as the above panels and added 68\% confidence intervals to a random selection of 50 points to show that the confidence in any one point is low; however, the bulk of the posterior is below the cut for many events. Furthermore, even events well above the line have tails that extend well into the region below the cut line. Thus, the numbers of point estimates below the line (107 and 283 for OGLE-III and IV on the left and right, respectively) are likely conservative.}\label{fig:fig18}
\end{figure*}

For either type of black hole, as the mass increases the event moves down and to the right in $t_E-\pi_E$ space, meaning searching for longer time scale events with less microlensing parallax is the best way to identify black hole lenses photometrically. This has not been fully appreciated in any black hole search to date; although, \citet{2020ApJ...889...31L} pointed it out in their Figure 13. Estimating by eye from their Figure 13, the purity of astrophysical black holes is very high with events satisfying:
\begin{equation}
    \log_{10}\pi_E < 0.8\,\log_{10}t_E + b, 
\end{equation}where b is an arbitrary intercept. We varied the intercept of this inequality until the mock OGLE-III and IV catalogs (separately) from \citet{2020ApJ...889...31L} yielded a 50$\%$ purity of black holes. We found $b=-2.67$ for OGLE-III and  $b=-2.57$ for OGLE-IV. We then checked our fits for these two surveys. Using the median marginal $\log_{10}t_E$ and $\log_{10}\pi_E$ values, 107 and 283 events lie within the region of 50$\%$ purity (see Figure \ref{fig:fig18}). 

We note that this does not mean that there are necessarily 50\% of 107 and 283 black holes detected in OGLE-III and IV, respectively. In fact, there are likely more. The systematics involved with these data (seasonal gaps in the observability from one hemisphere, noisy ground based data, time-domain variability of the baseline, unresolved blended stars that are not well characterized, to name a few) make it challenging to estimate small $\pi_E$ for many black hole events. It is likely that $\pi_E$ values from our modeling are over-estimated for many low SNR events with intrinsically small $\pi_E$. In those cases, the posterior is more heavily weighted by the prior PDF than the likelihood (the data) and thus, the median value of the $\pi_E$ marginal posterior samples is skewed high. Our prior was developed from the \textsf{PopSyCLE} simulations and is strongly driven by stellar microlensing events with higher $\pi_E$ than the typical black hole microlensing event toward the Galactic bulge. This is also evident in Table \ref{tab:injection} where the mean of $\log_{10} \pi_E$ is biased high -- when the sampler can identify the parallax signal, it fits it well. This occurs more frequently for intrinsically high $\pi_E$ microlensing events, but when it cannot fit for $\pi_E$ well, it reverts to the prior. This is especially evident in Figure \ref{fig:fig18}, for $t_E<30$ days, where the parallax signal is small no matter the intrinsic value. This is the cause of the ridge of events that align with $\pi_E\approx 0.1$. Since we are plotting median values, we are effectively reporting the median of the prior for these events.

This, coupled with the fact that the black hole purity increases as $t_E$ increases and $\pi_E$ decreases and there are a significant number of events offset in this direction from the 50\% purity line means that 50\% of 107 and 283 black holes detected in OGLE-III and IV is likely a lower limit based on the \citet{2020ApJ...889...31L} model. This is supported by the lower panels in Figure \ref{fig:fig18}, where the posteriors for individual events near the cut line clearly extend to both sides. This is the expectation though, the mass-distance degeneracy in photometric microlensing data typical makes inferring individual events as black holes to be very challenging. We have presented this figure to motivate population studies of photometric microlensing. Further estimation on our samples is beyond the scope of our analysis here, but it will be explored in detail in a follow-on paper (Dawson et al. in preparation). 

Another avenue for selecting black hole candidates is to make cuts on $b_{\text{sff}}$ \citep[e.g.,][]{2016MNRAS.458.3012W}. However, this parameter is notoriously challenging to fit with OGLE data as the vast majority of events are highly blended, and it is difficult to find events where all of the flux is from the source star. \citet{2020AA...636A..20W} have shown that when Gaia parallax and proper motion can be coupled with the microlensing fits for select events with high $b_{\text{sff}}$, this complexity can be reduced, enabling this potentially powerful method. 

Even without blending cuts, a relatively pure set of candidates could be generated to follow up with expensive astrometric observations to model the astrometric shift in the photometric centroid. We are actively carrying out such a survey with Keck adaptive optics (PI Lu).

\subsection{Conclusions}
We have modeled all public Galactic bulge microlensing events from OGLE-III and IV, yielding the most comprehensive public catalog MCMC fits to microlensing light curves to date. Our model includes both microlensing parallax due to Earth's orbital motion and a GP model for variable baseline flux due to intrinsic stellar variability. This was a computationally intensive task, and our modeling used roughly one million CPU-hours on Lawrence Livermore National Laboratory's high-performance computing resources\footnote{We used the Quartz computing system: \url{https://hpc.llnl.gov/hardware/platforms/Quartz}.}. Below we offer a list of the main takeaways of our analysis. 

\begin{itemize}
    \item We obtained 3560 from \citet{2015ApJS..216...12W} and 5790 microlensing events from \citet{2019ApJS..244...29M}.
    
    \item Our PSPL$+$parallax$+$GP model was able to capture and separate the yearly parallax signal and the intrinsic variability in the baseline from flux of the source and/or blended light. This represents the largest ever catalog of microlensing events modeled the parallax signal due to Earth's motion around the Sun. 
    
    \item We have made public 20,000 samples for all 12 parameters in our model (7 microlensing and 5 GP) for all 9,350 OGLE-III and IV events we analyzed. 
    
    \item Non-identifiability is a common critique of GP analyses. We completed an analysis for bias in our results and explained our findings in terms of our model. We did not identify any issues with the GP model systematically over-fitting the parallax signal that should have been modeled by the microlensing model.
    
    \item We developed a Bayesian hierarchical forward model for the $t_E$ distribution following the sample re-weighting method developed by \citet{2010ApJ...725.2166H}. We found that the parallax and GP modeling favors fewer events with $t_E>100$ days than were reported by \cite{2015ApJS..216...12W,2019ApJS..244...29M}. Motivated from theoretical modeling \citep{1996ApJ...473...57M}, we showed how different physical model choices can manifest starkly different $t_E$ distributions (Figure \ref{fig:fig17}). One could follow our population analysis method and use our published samples to forward model physical characteristics of the galaxy. 
    
    \item Using \textsf{PopSyCLE} \citep{2020ApJ...889...31L}, we demonstrate how relatively pure catalogs of black holes can be identified from photometry alone. We found 390 microlensing events in OGLE-III and IV that \textsf{PopSyCLE} simulations suggest are $\sim50\%$ black hole lensing events. While this benefits from the full rise and fall of the microlensing event, if parallax can be constrained below such a cut on the rising portion of an event, this technique could be used to identify the best targets for astrometric follow up.
    
    \item We recommend not applying long $t_E$ selection limits in future population studies of microlensing events, as they greatly impact the ability to conduct black hole studies.
    
    \item At a minimum, when working with OGLE light curves, the baseline flux must be checked for quasi-periodic and low-order polynomial trends, especially for long time-scale events. We demonstrated catastrophic failures in an MCMC analysis by not modeling these parameters. We recommend the use of additional model parameters to specifically handle these systematics where present.
\end{itemize}

\acknowledgments
We thank the Optical Gravitational Lensing Experiment (OGLE) collaboration for their decades long microlensing survey and for making many of their microlensing event light curves publicly available. The OGLE project received funding from the National Science Centre, Poland, grant MAESTRO 2014/14/ A/ST9/00121 to A.U.

We thank Peter McGill, Amber Malpas, and Dan Foreman-Mackey for fruitful brainstorming and discussion at the 2019 Microlensing meeting in New York. We also thank Tim Axelrod, Amanda Muyskens, Peter Nugent, Josh Meyers, and Bob Armstrong for useful discussions during this project. 

we and was supported by the LLNL-LDRD Program under Project 17-ERD-120. C.Y.L. and J.R.L. acknowledge support by the National Aeronautics and Space Administration (NASA) under contract No. NNG16PJ26C issued through the WFIRST Science Investigation Teams Program. M.S.M acknowledges support from the University of California Office of the President for the UC Laboratory Fees Research Program In-Residence Graduate Fellowship (Grant ID: LGF-19-600357). F.B. acknowledges financial support from the Science and Technology Facilities Council (STFC) in the UK for his PhD studentship. 

\software{This research has used NASA's Astrophysics Data System Bibliographic Services. This research used \textsf{exoplanet} \citep{exoplanet} and its dependencies \citep{exoplanet:agol19, exoplanet:astropy13, exoplanet:astropy18,exoplanet:exoplanet, exoplanet:kipping13, exoplanet:luger18, exoplanet:pymc3,exoplanet:theano}, as well as \textsf{celerite} \citep{2017AJ....154..220F}. This research used \textsf{NumPy} \citep{van2011numpy} and \textsf{Astropy}, a community-developed core Python package for Astronomy \citep{2018AJ....156..123A, 2013AA...558A..33A}. This research used matplotlib, a Python library for publication quality graphics \citep{Hunter:2007}. Finally, we used \textsf{corner.py} \citep{corner}.}

\bibliography{ogle}
\bibliographystyle{aasjournal}
\end{document}